\documentclass[usegraphicx, usenatbib]{mn2e}

\newcommand{\aap}   {A\&A}
\newcommand{\aj}    {AJ}
\newcommand{\apj}   {ApJ}
\newcommand{\apjl}  {ApJ}
\newcommand{\araa}  {ARA\&A}
\newcommand{\mnras} {MNRAS}
\newcommand{\nat}   {Nature}

\newcommand{\pasp}  {PASP}

\newcommand{\tmax}{\tau^{\rm max}}
\newcommand{\tint}{\tau_{\rm i}}
\newcommand{\tacc}{\tau_{\rm a}}
\newcommand{\tmag}{\tau_{\rm m}}

\newcommand{\mdotacc}{\dot M_{\rm a}}
\newcommand{\gammacrit}{\gamma_{\rm c}}

\newcommand{\rco}{R_{\rm co}}
\newcommand{\rt}{R_{\rm t}}
\newcommand{\rin}{R_{\rm in}}
\newcommand{\rout}{R_{\rm out}}

\newcommand{\omegaeq}{\Omega_*^{\rm eq}}

\title[The spin of accreting stars]{The spin of accreting stars: dependence on \\ magnetic coupling to the disc}

\author[Matt \& Pudritz]{Sean Matt\thanks{CITA National Fellow} and
Ralph E. Pudritz \\Physics \& Astronomy Department, McMaster
University, Hamilton ON, Canada L8S 4M1 \\ matt@physics.mcmaster.ca,
pudritz@physics.mcmaster.ca}

\begin{document}


\pagerange{\pageref{firstpage}--\pageref{lastpage}} \pubyear{2004}

\maketitle


\label{firstpage}

\begin{abstract}

We formulate a general, steady-state model for the torque on a
magnetized star from a surrounding accretion disc.  For the first
time, we include the opening of dipolar magnetic field lines due to
the differential rotation between the star and disc, so the magnetic
topology then depends on the strength of the magnetic coupling to the
disc.  This coupling is determined by the effective slip rate of
magnetic field lines that penetrate the diffusive disc.  Stronger
coupling (i.e., lower slip rate) leads to a more open topology and
thus to a weaker magnetic torque on the star from the disc.  In the
expected strong coupling regime, we find that the spin-down torque on
the star is more than an order of magnitude smaller than calculated by
previous models.  We also use our general approach to examine the
equilibrium (`disc-locked') state, in which the net torque on the star
is zero.  In this state, we show that the stellar spin rate is roughly
an order of magnitude faster than predicted by previous models.  This
challenges the idea that slowly-rotating, accreting protostars are
disc locked.  Furthermore, when the field is sufficiently open (e.g.,
for mass accretion rates $\ga 5 \times 10^{-9} M_\odot$ yr$^{-1}$, for
typical accreting protostars), the star will receive no magnetic
spin-down torque from the disc at all.  We therefore conclude that
protostars must experience a spin-down torque from a source that has
not yet been considered in the star-disc torque models---possibly from
a stellar wind along the open field lines.

\end{abstract}


\section{Introduction} \label{sec_intro}

Accretion discs are responsible for some of the most energetic and
spectacular phenomena in many classes of astrophysical objects,
including protostars, white dwarfs (cataclysmic variables and
intermediate polars), neutron stars (binary X-ray pulsars), and black
holes (both stellar mass X-ray transients and supermassive active
galactic nuclei).  Gravitational potential energy liberated by the
accretion process gives rise to exceptional luminosity excesses and
can drive powerful jets and outflows.  Accretion onto the central
object can occur only as quickly as angular momentum can be
transported away from the system.  Furthermore, the accretion of disc
material, which has high specific angular momentum, spins up the
central object, if the object rotates at less than the break-up rate.
It is therefore surprising that the central objects (hereafter
`stars') are often observed to spin far below their break-up rates, in
spite of long-lived accretion.  Why does this happen?

There is good evidence that accretion onto magnetized stars occurs
along closed magnetospheric field lines that connect the star to the
inner edge of the disc.  For instance, theoretical models of this sort
have been successful in explaining numerous observed features in
accreting protostars \citep[e.g.,][]{hayashi3ea96, goodson3ea99,
muzerolle3ea01}, intermediate polars \citep[e.g.,][]{patterson94}, and
X-ray pulsars \citep[e.g.,][]{jossrappaport84, alykuijpers90,
katoea01, kato3ea04}.  In some cases, there is even direct evidence
that the stars are magnetized, namely for accreting protostars
\citep{johnskrullea99}, intermediate polars \citep{piirola3ea93}, and
X-ray pulsars \citep{makishimaea99}.

Magnetic fields can also be effective at transferring angular momentum
away from the star, possibly explaining the observed rotation rates.
Torques on the star that are exerted by magnetic field lines anchored
to the star and that are also connected to the disc have been
calculated by several authors (e.g., \citealp{ghoshlamb79b};
\citealp{cameroncampbell93}; \citealp{lovelace3ea95};
\citealp{wang95}; \citealp{yi95}; \citealp[][hereafter
AC96]{armitageclarke96}; \citealp{rappaport3ea04}).  Under certain
circumstances, this torque can counteract the angular momentum
deposited by accretion, leading to a net spin-down of the star
(possibly explaining spin-down episodes observed in X-ray pulsars;
\citealp{ghoshlamb78}; \citealp{lovelace3ea95}) or giving rise to an
equilibrium state, in which the net torque on the star is zero
(possibly explaining the slow spin of some accreting protostars;
\citealp[][hereafter K91]{konigl91}).  In this equilibrium state, the
spin rate of the central object depends on the accretion rate in the
disc, and so a system is then considered to be `disc locked.'  Since
these models for the magnetic star-disc interaction show that
accreting stars can spin more slowly than the break-up rate, there is
a general perception that the presence of an accretion disc in any
system leads to slow rotation rates.  This idea of disc locking has
been applied to a variety of problems.  As an example, in systems
where the moment of inertia of the star is changing (e.g., during
contraction), some authors have assumed that disc-locking keeps the
star at a constant spin rate (e.g., as applied to protostars by
\citealp{bouvier3ea97, sills3ea00, barnes3ea01, tinker3ea02,
rebull3ea04}; and suggested for stellar collision products by
\citealp{leonardlivio95, sillsea01, demarcoea04}).

There is a nagging problem with this physical picture, however,
because the magnetic torque calculations discussed above (with the
exception of \citealp{lovelace3ea95}) assume that the stellar magnetic
field remains largely closed and that field lines connect to a large
portion of the disc\footnote{ Note that the X-wind model of
\citet[][and subsequent works]{shuea94} is unique among the star-disc
interaction theory in the literature, since it assumes that a system
will always accrete very near its disc-locked state. The magnetic
field geometry employed by the X-wind model is also unique and was
designed, in part, to avoid the problem of field opening due to
differential twisting, as considered in this paper.  Therefore, much
of our discussion does not apply to the X-wind.}.  This assumption is
questionable because closed magnetic structures tend to open when
enough energy is added to them, and these systems possess a natural
source of energy in the form of gravitational potential energy that is
released during disc accretion.  This energy release can drive
outflows and twist field lines, thereby adding energy to the magnetic
field.  Thus, the general surplus of energy in accreting systems
suggests that associated magnetic fields should be dominated by open,
rather than closed, topologies.  How are low spin rates achieved in
this case?

In this paper, we generalize the star-disc interaction model to
include the effect of varying field topology (i.e., connectedness).
We consider the mechanical energy that is added to the field via
differential rotation between the star and disc as the only mechanism
responsible for opening the field (though our formulation is easily
adaptable for other mechanisms).  The time-dependent behavior of a
dipole stellar field attached to a rotating, conducting disc has been
studied, using an analytic approach, by several authors (e.g.,
\citealp{lyndenbellboily94, agapitoupapaloizou00};
\citealp[][hereafter UKL]{uzdensky3ea02}).  They have shown that, as
the differential twist angle between the star and disc $\Delta \Phi$
monotonically increases, the torque exerted by field lines first
reaches a maximum value, then decreases.  This occurs because
azimuthal twisting of the dipole field lines generates an azimuthal
component to the field, and the magnetic pressure associated with this
component acts to inflate the field, which then balloons outward at an
angle of $\sim 60^\circ$ from the rotation axis, causally
disconnecting the star and disc \citep[see also][]{aly85,
alykuijpers90, newman3ea92, lovelace3ea95, bardouheyvaerts96}.
Typically, this inflation or opening of the field occurs when a
critical differential rotation angle of $\Delta \Phi_{\rm c} \approx
\pi$ has been reached, and the amount of flux that opens depends on
the strength of the magnetic coupling of the field to the disc (UKL).
This analytic work on the field opening has been corroborated by
time-dependent, numerical magnetohydrodynamic simulations of the
stellar dipole-disc interaction \citep[][]{hayashi3ea96, goodson3ea97,
goodson3ea99, millerstone97, katoea01, kato3ea04, mattea02,
romanovaea02, kuker3ea03}.

Our primary goal in this paper is to determine the effect of the
topology of the magnetic field on the torques in the steady-state,
star-disc interaction model.  In a previous paper
\citep{mattpudritz04a}, we gave a brief outline of this theory and
showed that a more open (i.e., less connected) field topology results
in a spin-down torque on the star that is less than for the closed
field assumption.  Consequently, the equilibrium state (with a net
zero torque) features a faster spin than predicted by previous models,
which calls into question the general belief that accretion discs
necessarily lead to slow rotation.  The present paper contains a more
detailed presentation of the theory and our assumptions, and we
consider all possible spin states of the system (not just the
equilibrium state).  We also extend our analysis to show that there
are at least three different modes in which a magnetic star-disc
system can operate.  Our analysis is applicable to all classes of
magnetized objects that accrete from Keplerian discs.  However, since
an abundance of observational data exists for accreting protostars, in
particular for classical T Tauri stars (CTTS's), we adopt a set of
fiducial parameters that are appropriate for these systems and discuss
various aspects of the model in this context.

Section \ref{sec_stardisc} contains a formulation of the general
model.  The special case of a disc locked system is the topic of
section \ref{sec_equilib}.  The final section (\S
\ref{sec_discussion}) contains a summary of our results and includes a
list of problems with using the disc-locking scenario to explain CTTS
spins, plus a discussion of three possible configurations of the
general system.

\section{Star-disc interaction model} \label{sec_stardisc}

Magnetic, star-disc interaction models in the literature differ in
their various assumptions, adopted parameter values, and in the
introduction of `fudge factors,' but they are quite similar on the
whole \citep[for a review, see][]{uzdensky04}.  We formulate a general
model that builds upon this previous work (mostly following AC96), by
including the effect of varying magnetic field topology, via the
introduction of the physical parameters $\beta$ and $\gammacrit$
(defined below).

According to the usual model assumptions, a rotating star is
surrounded by a thin, Keplerian accretion disc.  The angular momentum
vector of the disc is aligned with that of the star, which rotates as
a solid body and at a rate that is some fraction of break-up speed,
defined by
\begin{eqnarray}
\label{eqn_f}
f \equiv \Omega_* \sqrt{{R_*^{3}}\over{G M_*}},
\end{eqnarray}
where $\Omega_*$, $R_*$, and $M_*$ are the angular rotation rate,
radius, and mass of the star, respectively.  Note that $f$ is always
within the range from zero to one\footnote{ Disc accretion solutions
do exist for $f > 1$ \citep{pophamnarayan91, paczynski91}, in which
the star is actually spun down by accretion toward $f = 1$, even
without any magnetic torques.  However, we only consider cases with $f
\le 1$ in this paper, since this characterizes the spin of observed
protostellar systems.}.  The disc rotates at a different angular rate
than the star at all radii, except at the singular corotation radius
given by
\begin{equation}
\label{eqn_rco}
\rco = f^{-{2 / 3}} R_*.
\end{equation}
For $r < \rco$, the disc rotates faster than the star, while for $r >
\rco$, the angular rotation rate of the star is greater than that of
the disc.

The disc is assumed to be in a steady-state wherein the mass accretion
rate $\mdotacc$ is constant in time and at all radii.  In a real disc
from which winds are launched, $\mdotacc$ may have a weak radial
dependence, but we assume this has a negligible effect on the model.
A rotation-axis-aligned dipole magnetic field, anchored into the
stellar surface, also connects to the disc.  The field is strong
enough to truncate the disc at some inner location $\rt$ from where
disc material is subsequently channeled along magnetic field lines as
it accretes onto the star.  In general, the disc may have its own
magnetic field (either generated in a disc dynamo or carried in by the
disc from larger scales).  We do not specifically include this field
in the model, though it may be responsible for angular momentum
transport in the disc (providing $\mdotacc$) and may also aid in the
connection of the stellar field to the disc.  Within the disc, the
kinetic energy of the gas is much greater than the magnetic energy of
the stellar field, but the region above the star and disc (the corona)
is filled with low density material, and so the corona is magnetically
dominated.

In this configuration, the magnetic field connects the star and disc
by conveying torques between the two.  Torques are conveyed on an
Alfv\'en wave crossing time, which is much shorter than the Keplerian
orbital time.  Everywhere that the magnetic field connects the star to
the disc, exept at $\rco$, the magnetic field is twisted azimuthally
by differential rotation between the two.  Inside $\rco$ the field is
twisted such that field lines `lead' the stellar rotation, so torques
from field lines threading the region $r < \rco$ act to spin up the
star (and spin down the disc).  Conversely, torques from field lines
threading $r > \rco$ act to spin the star down (and spin the disc up).
The accretion of disc matter onto the star also deposits angular
momentum onto the star.

\begin{figure}
\centerline{\includegraphics[width=20pc]{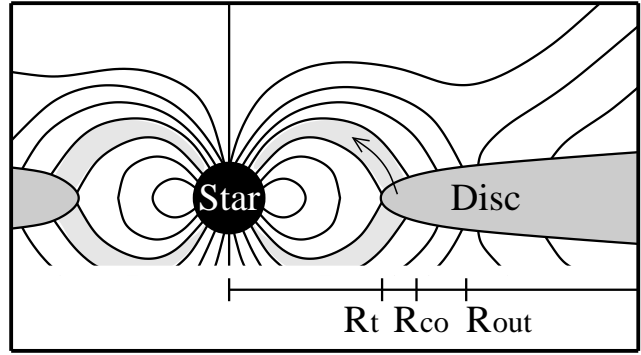}}

\caption{Magnetic star-disc interaction.  The stellar field connects
to a region of the disc, from $\rt$ to $\rout$, reaching beyond $\rco$
in this case.  The stellar field dominates the accretion flow onto the
star (arrow).
\label{fig_cartoon}}

\end{figure}

In order for disc material to accrete, $\rt$ must be less than $\rco$
so that accreting material loses angular momentum to the star as it
falls inward.  In order for the star with $f \le 1$ to feel any spin
down torques from the disc, the stellar field must connect to the disc
beyond $\rco$.  Under this condition, the field that connects outside
$\rco$ transfers angular momentum from the star to the disc.  To
maintain a steady accretion rate, the disc must then transport this
excess angular momentum outward, resulting in an altered disc
structure \citep{sunyaevshakura77a, sunyaevshakura77b, spruittaam93,
rappaport3ea04}.  We will define $\rout$ as the outermost radial
extent of the magnetic connection, and the usual assumption is that
$\rout \gg \rco$.  Figure \ref{fig_cartoon} illustrates the basic
picture, and shows the locations of $\rt$, $\rco$, and $\rout$ for a
possible configuration of the star-disc interaction model.

The assumption of a dipole field refers to the poloidal component of
the field, $B_{\rm p} = (B_r^2 + B_z^2)^{1/2}$, where $B_r$ and $B_z$
are the cylindrical $r$ and $z$ components of the magnetic field, and
the closed field only exists in the region interior to $\rout$.  The
dipole field is used for simplicity and because it has the weakest
radial dependence ($B_z \propto r^{-3}$ along the equator) of any
natural magnetic multipole.  In reality, the twisting of dipole field
lines alters the poloidal field, but this perturbation should be
slight in the region where the magnetic field remains closed (as
justified by the work cited in \S \ref{sec_intro}, e.g., UKL, and see
also \citealp{liviopringle92}).

For discussion throughout this paper, it is often instructive to use
physical units, especially for comparison with observations.  For this
purpose, we adopt a set of observationally determined `fiducial
parameters' that are appropriate for CTTS's \citep[e.g.,
see][]{johnskrullgafford02}:
\begin{eqnarray}
\mdotacc &=& 5 \times 10^{-8} \; M_\odot \; {\rm yr}^{-1},  \nonumber \\
M_* &=& 0.5 \; M_\odot,  \nonumber \\
R_* &=& 2 \; R_\odot,  \nonumber \\
B_* &=& 2 \times 10^3 \; {\rm G},  \nonumber
\end{eqnarray}
where $B_*$ is the stellar magnetic field strength at the
equator\footnote{ These fiducial parameters are slightly different
than used for figures 2 and 3 of \citet{mattpudritz04a}, who
considered the specific case of the CTTS BP Tau.}.  However, our
formulation of the problem is applicable to any magnetic star-disc
system.


     \subsection{Twisting and slipping of magnetic field lines} 
     \label{sec_twist}

The torque exerted by magnetic field lines threading an annulus of the
disc of radial width $d r$ is given by (e.g., AC96)
\begin{eqnarray}
\label{eqn_dtmag}
d \tmag = -\gamma {\mu^2 \over r^4} d r,
\end{eqnarray}
where
\begin{eqnarray}
\gamma \equiv B_\phi / B_z.
\end{eqnarray}
Here, $\mu$ is the dipole moment and $B_\phi$ is the azimuthal
component of the magnetic field.  The radial component, $B_r$, is
assumed to be negligible within the disc, the torque has been
vertically integrated through the disc, and $B_\phi$ refers to the
value at the disc surface.  Here, and throughout this paper, we choose
the sign of the torque to be relative to the star such that a positive
torque spins the star up, and consequently spins the disc down (and
vice versa for a negative torque).

The differential magnetic torque of equation \ref{eqn_dtmag} depends
not only on $\mu$, but also $\gamma$, which is the `twist\footnote{
This should not be confused with the `twist angle' of the footpoints
of the field, $\Delta \Phi$, as discussed by UKL, though $\Delta \Phi$
and $\gamma$ are intimately related.},' or pitch angle, of the field.
The total (integrated) magnetic torque will also depend on the size
and radial location of the magnetically connected region in the disc.
While $\mu$ is a constant parameter of the system, the radial
dependence of $\gamma$ depends on the physical coupling of the
magnetic field to the disc.

In general, the coupling is not perfect.  Magnetic forces act to
resist the twisting of the field, and so the field will `slip'
backward through the disc at some rate $v_{\rm d}$ proportional to
$\gamma$.  The exact slipping rate depends upon which physical
mechanism is at work.  In the literature, there are generally three
mechanisms discussed \citep[e.g., see][]{wang95}: 1) magnetic
reconnection in the disc, 2) reconnection outside the disc, and 3)
turbulent diffusion of the magnetic field through the disc.  We adopt
the latter mechanism, but we further discuss the other two, below.

The magnetic field slips azimuthally at a speed (e.g.,
\citealp{lovelace3ea95}; UKL)
\begin{eqnarray}
\label{eqn_vd}
v_{\rm d} = {\eta_{\rm t} \over h} \gamma = \beta v_{\rm k} \gamma,
\end{eqnarray}
where $\eta_{\rm t}$ is the turbulent magnetic diffusivity, $h$ is the
local disc scale height, $v_{\rm k}$ is the Keplerian orbital speed,
and we have introduced the dimensionless `diffusion parameter'
$\beta \equiv \eta_{\rm t} (h v_{\rm k})^{-1}$.  The variable $\beta$
simply parametrizes the coupling of the stellar magnetic field to the
disc such that $\beta \gg 1$ corresponds to weak coupling, and $\beta
\ll 1$ to strong coupling.

Generally, $\beta$ is a scale factor that compares $v_{\rm d}$ to
$v_{\rm k}$, and we have chosen this generic formulation so that the
system behavior is largely independent of any particular disc
model---as long as the disc obeys Keplerian rotation and provides a
steady accretion rate.  However, if we temporarily consider a standard
$\alpha$-disc \citep{shakurasunyaev73}, we may rewrite our diffusivity
parameter $\beta$ in a more physically revealing way:
\begin{eqnarray}
\beta = {\alpha \over P_{\rm t}} {h \over r},
\end{eqnarray}
where $\alpha$ has its usual meaning and $P_{\rm t}$ is the turbulent
Prandtl number, equal to the turbulent viscosity divided by $\eta_{\rm
t}$.  The disc turbulence is likely to be driven by the
magneto-rotational instability \citep[MRI;][]{balbushawley91}, which
follows the general behavior of an $\alpha$-disc.  Since both $\alpha$
and $h/r$ typically have weak radial dependences, and the value of
$P_{\rm t}$ is unknown, we assume that $\beta$ is constant in the
region of the disc connected to the stellar field.

The value of $\beta$ is not well constrained (AC96 used $\beta = 1$),
but extreme $\alpha$-disc parameters give an upper limit of $\beta \le
1$.  For a more reasonable estimate, note that a thin disc usually
means $h/r \la 0.1$, and $\alpha$ is typically in the range of 0.001
to 0.1 \citep{sanoea04}.  So, assuming $P_{\rm t}$ is of order unity,
$\beta \la 10^{-2}$.  We get a similar estimate using equation
\ref{eqn_vd} and reasonable guesses for CTTS disc parameters,
\begin{eqnarray}
\beta \approx 10^{-2} 
  \left({{\eta_{\rm t} \over 10^{16} \; {\rm cm}^2 \; {\rm s}^{-1}}}\right)
  \left({{h \over R_\odot}}\right)^{-1}
  \left({{v_{\rm k} \over 100 \; {\rm km} \; {\rm s}^{-1}}}\right)^{-1}.
\end{eqnarray}
However, given the uncertainties and possible variation among
different systems, and to assess the effect of the coupling of the
field to the disc, we retain $\beta$ as a free parameter.

\begin{figure}
\centerline{\includegraphics[width=22pc]{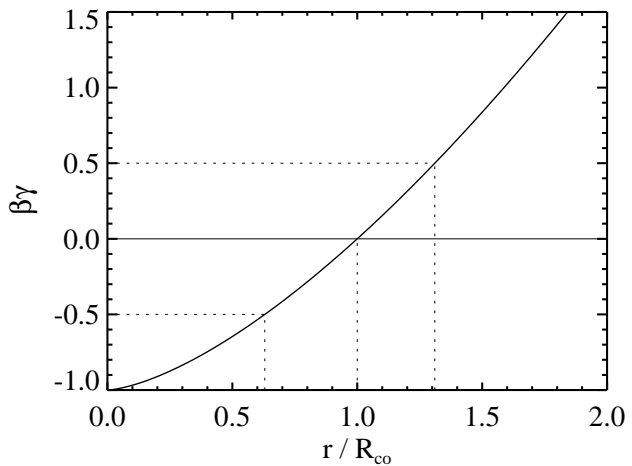}}

\caption{The magnetic twist ($\gamma \equiv B_\phi/B_z$) times the
diffusion parameter $\beta$, as a function of radius along the surface
of the disc (thick, solid line).  The dotted lines indicate critical
radii for a field-opening twist of $\gammacrit = \pm 0.5/\beta$
(chosen for illustrative purposes). The locations of $\rin$, $\rco$,
and $\rout$ are where $\gamma$ equals $-\gammacrit$, 0, and
$\gammacrit$, respectively.
\label{fig_gamma}}

\end{figure}

In the disc connected region, if $v_{\rm d}$ is anywhere less
(greater) than the local differential rotation speed between the star
and disc, $\gamma$ will increase (decrease) on an orbital
time-scale. Thus the magnetic field will quickly achieve a steady-state
configuration in which $v_{\rm d}$ equals the local differential
rotation rate (e.g., UKL), which gives
\begin{eqnarray}
\label{eqn_gamma}
\gamma = \beta^{-1} 
  \left[{(r / \rco)^{3 / 2} - 1}\right].
\end{eqnarray}
The solid line in Figure \ref{fig_gamma} shows the quantity $\beta
\gamma$ as a function of radius (normalized to $\rco$) along the
surface of the disc.  The magnetic twist is zero at $\rco$, and is
oppositely directed on either side of $\rco$.  Also, at a given
radius, the twist will be larger for smaller values of $\beta$ (and
vice versa).

We have assumed that the field coupling is determined by turbulent
diffusion, and when $\beta$ is constant, we find that $\gamma \propto
r^{1.5}$ (eq.\ \ref{eqn_gamma}, for $r \gg \rco$).  Other coupling
mechanisms (as discussed above) result in a different radial
dependence of $\gamma$.  For example, \citet{wang95} showed that if
the twist is limited by reconnection in the disc, $\gamma \propto
r^{1.6}$ (for $r \gg \rco$), while for reconnection outside the disc,
$\gamma$ approaches a constant value (for $r \gg \rco$).  On the other
hand, \citet{liviopringle92} and AC96, assumed $\gamma$ was limited by
reconnection in the stellar corona, and they used the same formulation
as equation \ref{eqn_gamma} (with $\beta = 1$).  In any case, note
that the radial dependence of the differential magnetic torque is
dominated by the falloff of the dipole magnetic field (which results
in the $r^{-4}$ dependence of eq.\ \ref{eqn_dtmag}).  Therefore, the
choice of magnetic coupling mechansim will not much influence our
results (AC96).  Similarly, a small radial dependence of $\beta$
(which we take as constant) will not introduce a large error.


          \subsubsection{Maximum twist for dipole field} \label{sub_maxtwist}

As discussed in section \ref{sec_intro}, several authors have shown
that dipole field lines will transition from a closed to an open
topology when a critical differential rotation angle of $\Delta
\Phi_{\rm c} \approx \pi$ has been reached.  This corresponds to a
critical field twist of $\gammacrit \approx 1$.  Since the twisting of
field lines does not significantly alter the poloidal configuration of
the field lines that remain closed, and as an approximation, we will
assume that the opening of field lines is only important for the
determination of the size of the connected region (i.e., to determine
$\rout$).  Thus, we include the effect of field line opening in the
steady-state torque theory in the following manner: we will use
equation \ref{eqn_gamma} only where $\gamma < \gammacrit$ and assume
that the field will be open everywhere else (a similar approach was
used by \citealp{lovelace3ea95} and justified by the work of UKL).  In
other words, wherever equation \ref{eqn_gamma} predicts $\gamma \ge
\gammacrit$, the magnetic connection is assumed to be severed, so the
star and disc are causally diconnected, such that no torques can be
conveyed between the two.  The size of the connected region in a
Keplerian disc is then limited to a finite radial extent near $\rco$,
where the differential rotation between the star and disc is the
smallest.

Will field lines, once opened by differential rotation, remain open?
It has been suggested that such field lines could reconnect in the
current sheet formed during the opening \citep{alykuijpers90,
uzdensky3ea02b}.  In order for this to be important, the time-scale
for reconnection should be comparable to or shorter than that for
field line opening.  It is not clear whether this is the case in these
systems \citep{uzdensky3ea02b, mattea02}, but even if it is, there are
other considerations.  First, due to the topology of the field,
reconnection must initially occur between open field lines at the
smallest radii (connecting to the lowest latitude on the star).  It is
possible that, if reconnection does occur, only a small amount of flux
will be able to reconnect before this newly connected field again
begins to open \citep[as in the simulations of][and see
\citealp{uzdensky3ea02b}]{goodson3ea99}.  In this case, the size of
the connected region (in a time-averaged sense) will be only slightly
larger than if the reconnection were never to occur.  Second, the
configuration of the opened field is favorable to launch
magnetocentrifugal flows from the disc \citep[][]{blandfordpayne82}.
We ignore such outflows in our model, but in a real system, they could
help to maintain an open magnetic field configuration.  Thus we
conclude that, once the field has opened, reconnection along the
current sheet is unlikely to significantly affect the size of the
connected region.

          \subsubsection{Maximal spin-down torque for maximal twist 
	    \label{sub_tmaxtwist}}

It is instructive to look at the maximum possible spin-down torque in
this system.  Regardless of any disc model or any magnetic coupling
physics, the largest possible magnetic torque on a star that connects
to a disc via a dipole magnetic field occurs when the field is
maximally twisted ($\gamma = \gammacrit$) at all radii along the
surface of the disc.  Spin-down torques on the star only occur along
field lines threading the disc outside $\rco$.  Also, the accretion of
mass from a Keplerian disc always adds angular momentum to the star.
Therefore, the largest net spin-down torque on the star occurs when
the disc is truncated exactly at the corotation radius ($\rt = \rco$),
the field threads the disc to $\rout \rightarrow \infty$, and the disc
does not accrete ($\mdotacc = 0$).  One then integrates equation
\ref{eqn_dtmag} from $\rco$ to $\infty$ to get
\begin{eqnarray}
\label{eqn_tmaxtwist}
\tmax_{\rm twist} = -{\gammacrit \over 3} {\mu^2 \over \rco^3}.
\end{eqnarray}
This is the absolute maximum spin-down torque that a star can undergo
from a disc that exists in the equatorial plane and to which the star
is connected via a dipole magnetic field.  It is even independent of
the rotation profile of the disc, except that angular rotation rate of
the disc is slower than the star outside some radius $\rco$.  It is
also independent of the angular momentum transport mechanism within
the disc.

To achieve this maximal torque requires that a) the field twist has no
radial dependence and b) the twist is very nearly equal to the maximum
allowed value of $\gammacrit$.  If the coupling of the field to a
Keplerian disc is determined by turbulent diffusion, a constant
$\gamma$ can only be achieved in the unlikely event that $\eta_{\rm
t}$ decreases with radius to exactly counteract the increase in
differential rotation rate.  Alternatively, reconnection in the
stellar or disc corona may also lead to a constant $\gamma$
\citep[][but see discussion in \S \ref{sub_maxtwist}]{alykuijpers90,
wang95}.  However, in either case, it is not clear why the value of
the constant twist would necessarily be near the maximal value
$\gammacrit$ (instead of, e.g., $0.1 \gammacrit$).  Though this torque
may not be very realistic, it is similar in strength to the spin-down
torque used in previous models (e.g., it is the equivalent to the
solution of AC96, for $\gammacrit = 1$).

          \subsubsection{Determination of $\rin$ and $\rout$}
	  \label{sub_rinrout}

To derive a more realistic magnetic torque, we adopt equation
\ref{eqn_gamma} for the radial dependence of $\gamma$.  Following
\citet{lovelace3ea95}, we assume that this equation is only valid
where $|\gamma| \le \gammacrit$, and that the field is open everywhere
else.  Thus, equation \ref{eqn_gamma} predicts that the outer radius
of the magnetically connected region in the disc is
\begin{eqnarray}
\label{eqn_rout}
\rout = (1 + \beta \gammacrit)^{2 / 3} \rco.
\end{eqnarray}
There is a corresponding location insided $\rco$ at which the twist
formally exceeds the critical value, given by
\begin{eqnarray}
\label{eqn_rin}
\rin = (1 - \beta \gammacrit)^{2 / 3} \rco.
\end{eqnarray}
The dotted lines in Figure \ref{fig_gamma} indicate these radii for
$\beta \gammacrit = 0.5$, in which case $\rin \approx 0.63 \rco$ and
$\rout \approx 1.31 \rco$.  It is evident that the field topology is a
function of $\beta \gammacrit$ such that more diffusion in the disc
allows for a larger connected region.  Note that, if $\beta \gammacrit
\ge 1$, the field can remain connected to the disc at any $r < \rco$
(since $\rin$ is then not defined).

The typical assumption of a closed magnetic topology, corresponding to
$\rout \rightarrow \infty$ \citep[e.g., AC96;][]{yi95}, is equivalent
to $\gammacrit \rightarrow \infty$---the field is allowed to twist to
arbitrarily large values without opening.  In order to consider the
effect of varying topology (i.e., where the field is open beyond some
finite $\rout$), we adopt a value of $\gammacrit = 1$ (as justified
by, e.g., UKL).  However, we will retain $\gammacrit$ as a parameter
in all of our formulae for a comparison between the two cases
($\gammacrit \rightarrow \infty$ and $\gammacrit = 1$) and so that
different values of $\gammacrit$ may be considered by the reader.  The
combined parameter $\beta \gammacrit$ appears throughout our
formulation.  We generally think of this parameter in two ways.
First, when $\gammacrit = \infty$, the stellar field is closed and
connects to the entire disc, and this represents the `standard'
star-disc interaction model.  Second, for the more realistic case of
$\gammacrit = 1$, the field topology is partially open, and $\rout$
then depends on $\beta$.

          \subsection{Three possible states of the system \label{sub_states}}

\begin{figure}
\centerline{\includegraphics[width=20pc]{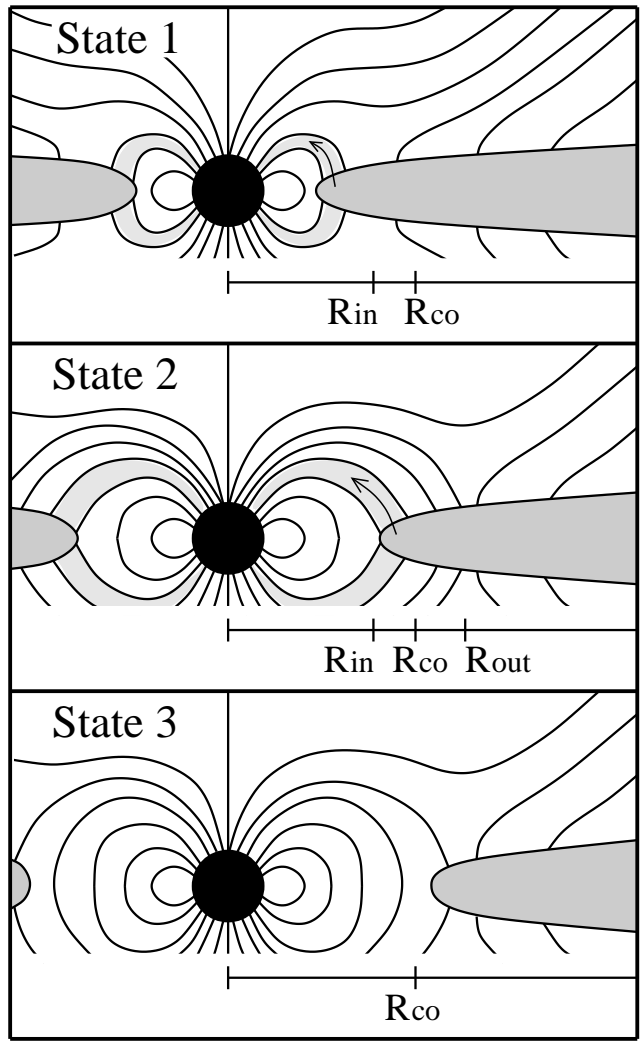}}

\caption{Three possible configurations of the magnetic star-disc
interaction.  There are two possible accreting states: either the
stellar field connects only to the inner edge of the disc (top panel)
or it connects to an extend region, reaching beyond $\rco$ (middle
panel).  In either case, the stellar field dominates the accretion
flow (arrows) onto the star.  Under certain conditions (e.g., low
$\mdotacc$), accretion onto the central star will cease, defining the
third, non-accreting state (bottom panel).
\label{fig_states}}

\end{figure}

The inner edge of the disc is delimited by $\rt$ (discussed in \S
\ref{sub_rt}), and so there are two possible magnetic configurations
of an accreting system, depending on the location of $\rt$ relative to
$\rin$.  First, if $\rt < \rin $, the stellar field will be largely
open, and equation \ref{eqn_gamma} is not valid anywhere.  The outer
radius of the magnetically connected region, $\rout^\prime$ (which is
not then determined by eq.\ \ref{eqn_rout}) will be near the inner
edge of the disc ($\rout^\prime \sim \rt$).  We will refer to this
situation as 'state 1' of the system.  In state 1, the star receives
no spin-down torques from the disc.

In `state 2,' $\rin < \rt < \rco$ and the star is magnetically
connected to the disc from $\rt$ to $\rout$.  State 2 represents the
typical configuration considered in many models and was discussed at
the beginning of this section.  Also, systems near their disc-locked
state (\S \ref{sec_equilib}) are always in state 2.

Finally, there exists a third possible, non-accreting state, 'state
3,' that occurs when the disc is overpowered by the magnetic field
(e.g., for fast rotation, large $\mu$, or small $\mdotacc$) and the
disc becomes truncated outside $\rco$ (e.g.,
\citealp{illarionovsunyaev75}).  In state 3, there are no positive
(spin-up) torques on the star, so it can never be in spin equilibrium
\citep{sunyaevshakura77b}.

Figure \ref{fig_states} illustrates the basic magnetic configuration
of each state.  One can think of this Figure, for example, as a
sequence (from top to bottom) of decreasing $\mdotacc$ (or increasing
$\mu$).  A steadily decreasing $\mdotacc$ may represent an
evolutionary sequence (e.g.) for protostars as one goes from class 0
sources to weak-lined T Tauri stars (class 3).  As $\mdotacc$
decreases from a system in state 1, the disc truncation radius (which
is at the inner edge of the disc in the Fig.)  moves outward and
eventually crosses the location of $\rin$ (entering state 2) and then
$\rco$ (state 3).  

The conditions for which a system transitions from an accreting state
to state 3 is unknown (see \citealp{rappaport3ea04} and \S
\ref{sub_state3}).  In the current work, we do not consider state 3,
other than to note that it occurs somewhere below the lower $\mdotacc$
limit of accreting systems.  Instead, we focus most of our attention
on state 2 and discuss state 1, where appropriate.

     \subsection{Torques between the star and disc \label{sub_torques}}

A combination of equations \ref{eqn_dtmag} and \ref{eqn_gamma} gives
the full radial dependence of the differential magnetic torque in the
system,
\begin{eqnarray}
\label{eqn_dtmagdr}
{d \tmag \over d r} = 
  {f^{8 / 3} \over \beta}
  {\mu^2 \over R_*^4}
  \left({r \over \rco}\right)^{-4}
  \left[{1 - \left({r \over \rco}\right)^{3 / 2}}\right].
\end{eqnarray}
where, for convenience, we have used equation \ref{eqn_rco} to express
the radius in units of $\rco$.  Furthermore, the angular momentum
carried by accreting material through each annulus of a Keplerian disc
(of width $d r$ and vertically integrated) equals $d \tacc = 0.5
\mdotacc (G M_* / r)^{-1/2} d r$ \citep[e.g.,][]{clarkeea95}.  This
can be combined with equation \ref{eqn_rco} to give the differential
accretion torque, as a function of $r/\rco$,
\begin{eqnarray}
\label{eqn_dtaccdr}
{d \tacc \over d r} =
  {1\over 2} \mdotacc f^{1 / 3}
  \left({G M_* \over R_*}\right)^{1 / 2}
  \left({r \over \rco}\right)^{-{1 / 2}}.
\end{eqnarray}

The assumption that $\mdotacc$ is constant at all radii in the disc,
requires that the net angular momentum transported away from each
annulus in the disc equals $d \tacc$.  The disc must therefore be
structured in such a way that the differential torques internal to the
disc, $d \tint$, satisfy
\begin{eqnarray}
\label{eqn_dtint}
d \tint \equiv d \tacc - d \tmag.
\end{eqnarray}
These internal torques could result from angular momentum transport
via (e.g.) turbulent viscosity \citep{shakurasunyaev73}, MRI
\citep{balbushawley91}, or disc winds \citep[see review
by][]{koniglpudritz00}.  If one assumes a particular angular momentum
transport mechanism in the disc, the solution to equation
\ref{eqn_dtint} determines the structure of the disc.  As an example,
for the case of $\alpha$ viscosity, \citet{rappaport3ea04} showed that
the disc can respond to external magnetic torques by increasing its
surface density in order to transport the additional angular momentum
outward.  A detailed treatment of the disc adjustment is not necessary
here, since we are presently concerned with torques on the star, and
we simply assume that the disc structures itself such that equation
\ref{eqn_dtint} is valid.

\begin{figure}
\centerline{\includegraphics[width=22pc]{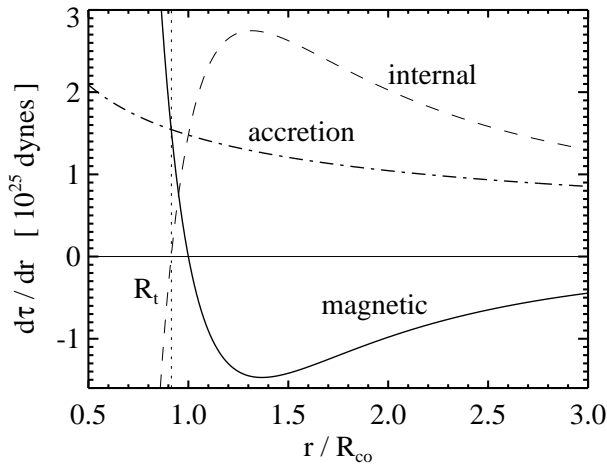}}

\caption{Differential torques in the fiducial CTTS system (see
discussion above \S \ref{sec_twist}) for $\beta = 1$ and $\gammacrit
\rightarrow \infty$, where the field is assumed to remain connected
over the entire disc.  The truncation radius, $\rt$, is where $d
\tmag = d \tacc$ and $d \tint = 0$, indicated by the
vertical dotted line ($\rt \approx 0.91 \rco$).  The system is shown
in its equilibrium state, where the net torque on the star is zero,
requiring a stellar spin period of 6.0 days ($\rco \approx 5.5 R_*$)
\label{fig_dtorques}}

\end{figure}

Figure \ref{fig_dtorques} shows the differential torques ($d \tau
/ d r$) as a function of $r / \rco$ for the adopted fiducial
parameters.  The solid, dash-dotted, and dashed lines in Figure
\ref{fig_dtorques} represent the differential torques from equations
\ref{eqn_dtmagdr}, \ref{eqn_dtaccdr}, and \ref{eqn_dtint},
respectively.  The system shown has $\beta = 1$ and $\gammacrit
\rightarrow \infty$, so that the field is connected to the entire
surface of the disc, and so that the Figure represents the closed
topology of several models in the literature (e.g., AC96).


The differential magnetic torque ($d \tmag$, solid line in Fig.\
\ref{fig_dtorques}) is strongest near the star, where the dipole field
is strongest, and it acts to spin up the star (and thus spins down the
disc) for $r < \rco$ .  At $\rco$, $d \tmag$ goes to zero, since
the field is not twisted ($\gamma = 0$) there.  Outside $\rco$, the
magnetic torque becomes stronger again, as the twist increases, though
now acting to spin the star down (and the disc up).  The dipole field
strength falls off faster with distance ($B_z \propto r^{-3}$) than the
magnetic twist increases ($\gamma \propto r^{3/2}$), so $d \tmag$
has a minimum value at $r/\rco \approx 1.37$ and then goes to zero as
$r \rightarrow \infty$.

The dashed line in the Figure ($d \tint$) gives us some information
about the disc structure.  In this case, the structure will be
significantly different than for a case with $d \tmag = 0$.  For a
very large $d \tmag$ outside $\rco$, the assumption that the disc can
counter act it (via an increase in $d \tint$) must eventually break
down, and the system would then be in state 3 (\S \ref{sub_states}).

          \subsubsection{Truncation radius} \label{sub_rt}

Inside the corotation radius, the stellar magnetic torque acts to
extract angular momentum from the disc, further enabling accretion
(not hindering accretion, as for $r > \rco$).  The differential
magnetic torque increases rapidly as one moves toward the star, and at
some point, $d \tmag = d \tacc$.  There, the external magnetic torque
alone is capable of maintaining $\mdotacc$.  Consequently, $d \tint$
goes to zero at the same radius and formally becomes negative for
smaller $r$ (see Fig.\ \ref{fig_dtorques}).  Negative $d \tint$ is
unrealistic, however, since it would require angular momentum transfer
inward through the disc, from slower spinning material to faster
spinning material, so the Keplerian disc does not exist where where $d
\tint \le 0$.  Thus, the disc truncation radius\footnote{ Note that
$\rt$ is related to the `fastness parameter,' $\omega$, in X-ray
pulsar literature: $\omega = (\rt / \rco)^{3/2}$.}, $\rt$, is where $d
\tint = 0$.

At $\rt$, the stellar magnetic field will quickly spin down the disc
material, forcing it into corotation with the star.  Sub-Keplerian
rotation leads to a free-fall of disc material onto the surface of the
star in a `funnel flow' along magnetic field lines (e.g., K91;
\citealp{romanovaea02}).  Whether or not the funnel flow originates
exactly from $\rt$ or from inside that radius is subject to debate
\citep[e.g.,][]{alykuijpers90}.  However, for the present discussion
of angular momentum transport, the most important thing is that $\rt$
defines the location where the stellar magnetic field completely
dominates over the disc internal stresses, and so all of the angular
momentum of disc material at $\rt$ will end up on the star.

By setting equation \ref{eqn_dtmagdr} equal to \ref{eqn_dtaccdr}, we
derive a relationship defining $\rt$,
\begin{eqnarray}
\label{eqn_rt}
\left({\rt \over \rco}\right)^{-7/2}
  \left[{1 - \left({\rt \over \rco}\right)^{3/2}}\right] =
  {\beta \over \psi} f^{-{7/3}},
\end{eqnarray}
where
\begin{eqnarray}
\label{eqn_psi}
\psi \equiv 2 \mu^2 \mdotacc^{-1} (G M_*)^{-1/2} R_*^{-7/2}
\end{eqnarray}
is a dimensionless parameter relating the strength of the magnetic
field to the strength of accretion.  This formula was also derived by
\citet{yi95}, but with different disc parameters in place of our
$\beta$.  For any given $\beta$, $f$, and $\psi$, there is only one
solution to equation \ref{eqn_rt} such that $\rt < \rco$.  A real
system may deviate slightly from our simple picture (e.g., of an
unperturbed dipole field), leading to an uncertainty in the exact
value of $\rt$.  However, due to the steep radial dependence of
$d \tmag$ relative to $d \tacc$, the location of $\rt$
should not be significantly affected.  For the system plotted in
Figure \ref{fig_dtorques}, the solution to equation \ref{eqn_rt} is
$\rt/\rco \approx 0.915$, represented by the vertical dotted line in
the Figure.

For a system in state 1 (with $\rt < \rin$), equation \ref{eqn_rt} is
not valid because the field will open (see \S \ref{sub_rinrout}).  A
substitution of $\rt < \rin$ in equation \ref{eqn_rt} indicates that
the system will be in state 1 if
\begin{eqnarray}
\label{eqn_trans}
f < (1 - \beta \gammacrit) (\gammacrit \psi)^{-3/7},
\end{eqnarray}
and it will be in state 2 for any larger $f$.  Note that condition
\ref{eqn_trans} can never be satisfied if $\beta \gammacrit \ge 1$
(since $\rin$ is then undefined), and so the system would then always
exist in state 2.  For the more likely case that $\beta \gammacrit \ll
1$, state 1 is a possible configuration of any system.

To determine $\rt$ in state 1, instead of using equation
\ref{eqn_dtmagdr} for the differential magnetic torque, one must
consider the maximum possible $d \tmag$ in order that the field
remains closed.  This is determined by using $\gamma = -\gammacrit$ in
equation \ref{eqn_dtmag}.  By setting this equal to equation
\ref{eqn_dtaccdr}, one finds
\begin{eqnarray}
\label{eqn_rt2}
\rt = (\gammacrit \psi)^{2/7} R_* = (2\gammacrit)^{2/7}
  (G M_*)^{-{1/7}} (\mdotacc)^{-{2/7}} \mu^{4/7}.
\end{eqnarray}
Note that this equation does not depend on the rotation rate or radius
of the star and has the same dependences on other system parameters as
in many previous theoretical works \citep[e.g.,][]{davidsonostriker73,
ghoshlamb79b, shuea94}.  Since $\rt < \rin$ in state 1, the field
lines will be open inside $\rco$.  Thus, in state 1, the stellar field
connects only to a small portion of the disc near $\rt$, from where a
funnel flow originates, and all exterior field lines are open, as
shown in the top panel of Figure \ref{fig_states}.  State 1 is
discussed further in section \ref{sub_3states}.

\begin{figure}
\centerline{\includegraphics[width=22pc]{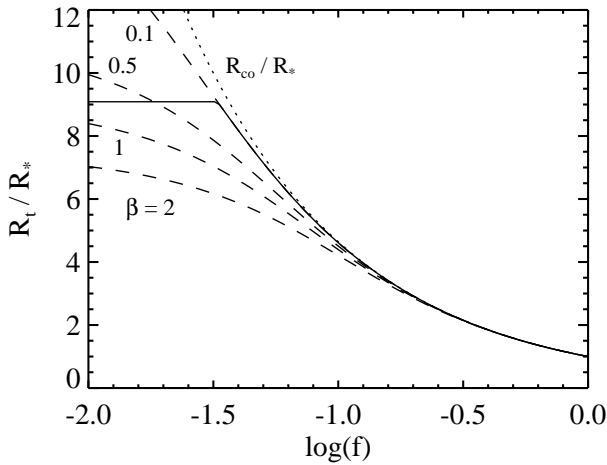}}

\caption{Predicted location of $\rt/R_*$ as a function of $\log(f)$,
for the fiducial CTTS system.  The dashed lines represent the
prediction from equation \ref{eqn_rt} for $\beta$ = 0.1, 0.5, 1.0, and
2.0, as indicated, and all have $\gammacrit \rightarrow \infty$ so
that the magnetic field is assumed to remain closed.  The solid line
is for $\beta = 0.1$, but with $\gammacrit = 1$, so that the field is
partially open.  The system with $\gammacrit = 1$ switches to state 1,
for $\log (f) \la -1.5$, and $\rt$ is then predicted by equation
\ref{eqn_rt2}.  The dotted line shows the location of $\rco$ (eq.\
\ref{eqn_rco}).
\label{fig_rtrstar}}

\end{figure}

Figure \ref{fig_rtrstar} shows the predicted location of $\rt$ in
units of $R_*$ for the adopted fiducial parameters, as a function of
the logarithm of the spin rate $f$.  For reference, the dotted line
shows the location of $\rco$ (eq.\ \ref{eqn_rco}).  The dashed lines
show $\rt$ for systems with $\gammacrit \rightarrow \infty$ and
$\beta$ = 0.1, 0.5, 1.0, and 2.0, as indicated in the Figure.  The
solid line shows the $\beta = 0.1$ case, but with a more realistic
value of $\gammacrit = 1$.  Note that $\rt$ is always less than
$\rco$, and that both decrease as $f$ increases.  For stronger field
coupling (smaller $\beta$), the field is more strongly twisted, and so
$\rt$ is closer to $\rco$.

All of the dashed lines in Figure \ref{fig_rtrstar} represent systems
with $\gammacrit \rightarrow \infty$, in which the field remains
closed for arbitrarily large magnetic twist.  Thus, these systems are
always in state 2, and the dashed lines are everywhere given by
equation \ref{eqn_rt}.  On the other hand, the solid line represents a
system with $\gammacrit = 1$, so it is in state 2 only when $f \ga
0.03$ (eq.\ \ref{eqn_trans}).  For smaller $f$, it is in state 1, and
$\rt$ is then determined by equation \ref{eqn_rt2}.  Since equation
\ref{eqn_rt2} is independent of $f$, state 1 is represented by the
constant value of $\rt \approx 9.1 R_*$ in the Figure.  Real systems
will likely have $\gammacrit = 1$, in which case the value of $\rt
\approx 9.1 R_*$ represents an upper limit for the fiducial CTTS
system, regardless $\beta$.

          \subsubsection{Accretion torque}

We assume that accreted disc material is quickly integrated into the
structure of the star and the accreted angular momentum is
redistributed into the stellar rotation profile.  So there is a torque
on the steadily accreting star that is given by $\tacc = \mdotacc
[l_{\rm d}(\rt) - l_*]$, where $l_{\rm d}(\rt)$ is the specific
angular momentum of the disc material at $\rt$ and $l_*$ is that of of
the star.  Combined with equation \ref{eqn_f}, and assuming solid body
rotation of the star, this becomes
\begin{eqnarray}
\label{eqn_tacc}
\tacc = \mdotacc (G M_* R_*)^{1/2} 
  \left[{(\rt / R_*)^{1/2} - k^2 f}\right],
\end{eqnarray}
where $k$ is the normalized radius of gyration ($k ^2 \approx 0.2$ for
a fully convective star; AC96).  This formula is valid for a system in
either state 1 or 2.

The term in square brackets in equation \ref{eqn_tacc} is
dimensionless and compares the accreting angular momentum (first term)
with how much the star already has (second term).  Note that the first
term will always be greater than or equal to one, while the second
term has a maximum value of $k^2$ (when $f = 1$).  Thus, the second
term is usually negligible (particularly when $f \ll 1$).

          \subsubsection{Magnetic torque} \label{sub_tmag}

When in state 2 (i.e., when $\rt > \rin$), the stellar field connects
to a significant portion of the disc, and one can integrate equation
\ref{eqn_dtmagdr} over the connected region, from $\rt$ to $\rout$, to
obtain the total magnetic torque on the star,
\begin{eqnarray}
\label{eqn_tmag1}
\tmag =  {1 \over 3 \beta}{\mu^2 \over \rco^3} 
  \left[{2 (\rco / \rout)^{3/2} - (\rco / \rout)^{3}}\right. \nonumber \\
  \left.{- 2(\rco / \rt)^{3/2} + (\rco / \rt)^{3}}\right].
\end{eqnarray}
This torque is independent of the detailed structure of the Keplerian
disc.  Also, equation \ref{eqn_tmag1} includes the dependence of the
magnetic torque on the field topolgy via the variable $\rout$.  For
example, the spin-down torque (found by setting $\rt = \rco$) exerted
by field lines connected out to $\rout \approx 2.4 \rco$ is one half
of the spin-down torque for $\rout \rightarrow \infty$.  Similarly,
for $\rout \approx 7.2 \rco$ or $\rout \approx 34 \rco$, the spin-down
torque is 90\% or 99\% (respectively) of the spin-down torque for
$\rout \rightarrow \infty$.  It is evident that, even when $\rout$ is
large, most of the spin-down torque comes from field lines connected
not too far from $\rco$.  This is simply because the differential
magnetic torque (eq.\ \ref{eqn_dtmagdr}) becomes very weak far from
the star.  Thus the typical assumption of $\rout \rightarrow \infty$
is not significantly effected by the fact that real discs have finite
radial extents, so long as they reach to several times $\rco$.


Above, we have taken $\rout$ as arbitrary, but our goal in this paper
is to consider the opening of the field from differential rotation, so
$\rout$ is then given by equation \ref{eqn_rout}, and the preferred
formulation of the magnetic torque becomes
\begin{eqnarray}
\label{eqn_tmag}
\tmag = {1 \over 3 \beta}{\mu^2 \over \rco^3} 
  \left[{2 (1 + \beta \gammacrit)^{-1} - (1 + \beta
  \gammacrit)^{-2}}\right. \nonumber \\ \left.{- 2(\rco / \rt)^{3/2} +
  (\rco / \rt)^{3}}\right].
\end{eqnarray}
This is exactly the solution found by AC96 for the special case of
$\beta = 1$ and $\gammacrit \rightarrow \infty$ (so that $\rout
\rightarrow \infty$), but our formulation includes the effect of field
opening via differential rotation, in which case, the field topology
depends on the diffusion parameter $\beta$.  The total magnetic torque
on the star can be either positive or negative, depending on the size
of the connected region inside $\rco$, compared to the connected
region oustide $\rco$.  In the next section \ref{sec_open}, we will
show that, for reasonable values of $\beta$ and $\gammacrit$, $\rout$
is very close to $\rco$, and the spin-down torque is significantly
effected.

     \subsection{Effect of opened field} \label{sec_open}

\begin{figure}
\centerline{\includegraphics[width=22pc]{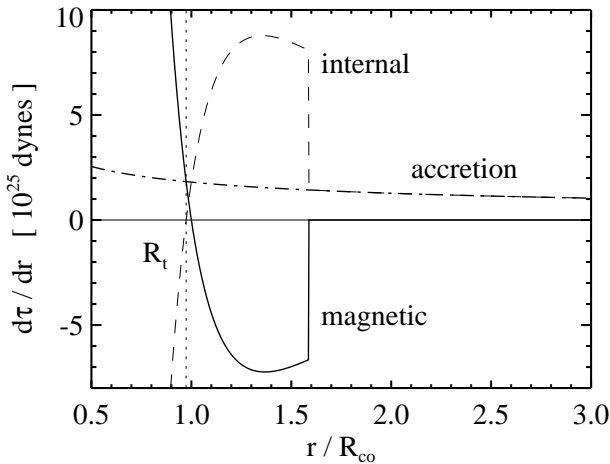}}

\caption{Same as Figure \ref{fig_dtorques}, except that the
equilibrium spin period is 3.3 days ($\rco \approx 3.7 R_*$), and
$\gammacrit = 1$, so that the magnetic field is not connected for $r >
\rout \approx 1.6 \rco$.  Also, $\rt \approx 0.97 \rco$.
\label{fig_dtorques2}}

\end{figure}

Figure \ref{fig_dtorques2} shows the differential torques in the
fiducial CTTS system with $\beta = 1$, as in Figure
\ref{fig_dtorques}.  However, unlike Figure \ref{fig_dtorques}, here
we show the system for $\gammacrit = 1$, so that the field is open
beyond $\rout \approx 1.59 \rco$.  The star now rotates significantly
faster, with a period of 3.3 days ($f \approx 0.14$), and $\rt \approx
0.974 \rco$.  A comparison between Figures \ref{fig_dtorques} and
\ref{fig_dtorques2} illustrates the effects of varying field topology
on the differential torques in the star-disc system.

There are some interesting differences between Figures
\ref{fig_dtorques} and \ref{fig_dtorques2}.  Most notably, the
differential magnetic torque in Figure \ref{fig_dtorques2} abruptly
goes to zero at the location of $\rout$, due to the assumption that
there is no torque on the star from the disc where the field lines are
open.  It is evident that a more open topology results in a smaller
connected region, which leads to a net (integrated) spin-down torque
that is smaller than for the completely closed topology.  Thus, a more
open topology results in a faster equilibrium spin rate, as can be
seen by comparing the stellar spin rate of 6.0 days for Figure
\ref{fig_dtorques} with 3.3 days for Figure \ref{fig_dtorques2}.

To quantify the effect of a more open topology on the magnetic torque
and to determine the dependence on $\beta$, we first consider only the
portion of the magnetic torque that acts to spin down the star by
setting $\rt = \rco$ in equation \ref{eqn_tmag}.  For normalization,
we use the maximum spin-down torque, $\tmax_{\rm twist}$ (eq.\
\ref{eqn_tmaxtwist}).  We define the ratio of the true spin-down
torque to this maximum torque as $\chi \equiv \tmag(\rt = \rco) /
\tmax_{\rm twist}$, which is given by
\begin{eqnarray}
\label{eqn_chi}
\chi = (\beta \gammacrit)^{-1}
  \left[{1 + (1 + \beta \gammacrit)^{-2}
  - 2 (1 + \beta \gammacrit)^{-1}}\right]
\end{eqnarray}
and only depends on the parameter $\beta \gammacrit$.  It is
immediately evident that for $\beta \gammacrit = 1$, $\chi = 1/4$, so
the spin-down torque is four times less than used by AC96, when one
considers a more realistic magnetic topology.

\begin{figure}
\centerline{\includegraphics[width=22pc]{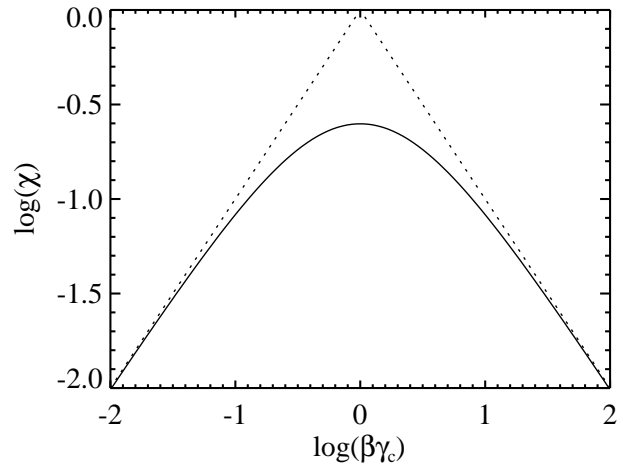}}

\caption{Logarithm of the ratio of real to maximal spin-down torque
(eq.\ \ref{eqn_chi}), as a function of $\log(\beta \gammacrit)$.
Dotted lines represent $\chi = \beta \gammacrit$ and $\chi = (\beta
\gammacrit)^{-1}$.
\label{fig_chi}}

\end{figure}

Figure \ref{fig_chi} illustrates the dependence of the torque ratio
$\chi$ on $\beta \gammacrit$.  It decreases as $\beta \gammacrit^{-1}$
for $\beta \gg 1$ and increases as $\beta \gammacrit$ for $\beta \ll
1$ (as revealed by Taylor expansion of eq.\ \ref{eqn_chi}).  The
limiting behavior of $\chi$ is indicated by the two dotted lines in
the Figure.  This behavior can be understood as a competition between
two effects: In the strong magnetic coupling limit ($\beta \ll 1$),
the field topology becomes more open for smaller $\beta \gammacrit$,
reducing the spin-down torque.  In the weak coupling limit ($\beta \gg
1$), the topology is largely closed, but the twisting of the field
lines is smaller for larger $\beta$, which reduces the differential
magnetic torque at all radii ($d \tmag \propto \beta^{-1}$).
These two effects conspire to give a maximal value of $\chi$ for the
special case of $\beta \gammacrit = 1$.  For the more likely value of
$\beta = 10^{-2}$, the spin-down torque is two orders of magnitude
lower than used by AC96.  While it may at first seem surprising that
strong magnetic coupling leads to weaker spin-down torques on the
star, further reflection reveals that this is necessarily true, since
stronger coupling leads to stronger twisting, which further
disconnects the star from the disc.

\begin{figure}
\centerline{\includegraphics[width=22pc]{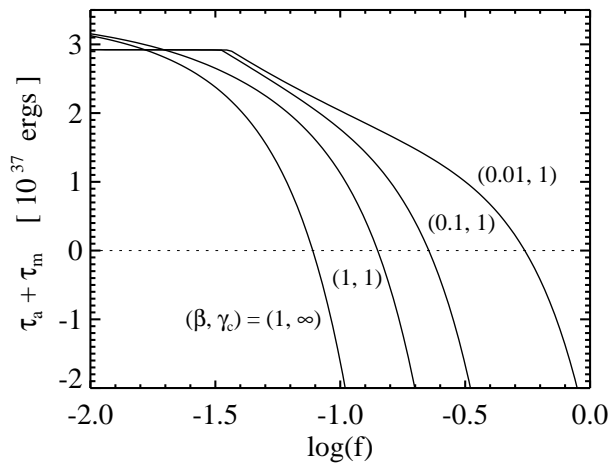}}

\caption{Net torque, $\tacc + \tmag$ (eqs.\ \ref{eqn_tacc} and
\ref{eqn_tmag}), on the star as a function of the logarithm of the
fractional spin rate, for the fiducial CTTS system. The four solid
lines represent $(\beta, \gammacrit) = (1, \infty)$, (1, 1), (0.1, 1),
and (0.01, 1), as indicated, corresponding to different field
topologies, ranging from completely closed to more open.  A system
will be in spin equilibrium when the net torque equals zero (dotted
line).
\label{fig_torques}}

\end{figure}

Finally, equations \ref{eqn_tacc} and \ref{eqn_tmag} can be used to
calculate the net torque on the star.  Figure \ref{fig_torques} shows
this net torque as a function of $\log(f)$ for the fiducial CTTS
parameters and for different values of $\beta$ and $\gammacrit$.  For
each case, we calcuate the net torque as follows: First, for a given
value of $\beta$ and $\gammacrit$, and for each value of $f$, we use
equation \ref{eqn_trans} to determine whether the system is in state 1
or 2.  We then find the location of the truncation radius, using
equation \ref{eqn_rt} if the system is in state 2, or equation
\ref{eqn_rt2} if in state 1.  Finally, we calculate the integrated
torques $\tacc$ (eq.\ \ref{eqn_tacc}) and $\tmag$ (eq.\
\ref{eqn_tmag}, if in state 2; $\tmag = 0$, if in state 1).  As
discussed in section \ref{sub_rt}, only cases with $\beta \gammacrit <
1$ can be in state 1, so only those cases show a transition at $\log f
\approx -1.45$ in Figure \ref{fig_torques}.  The Figure also shows the
effect of the field topology on the net torque on the star from the
disc.  It is evident that, when the magnetic field is partially open
($\gammacrit = 1$), the net torque is larger than for the case where
the field is everywhere closed ($\gammacrit = \infty$).  The spin rate
at which the net torque on the star is zero indicates the equilibrium
spin state, which is the topic of the next section.

\section{The disc-locked state} \label{sec_equilib}

The general theory presented in section \ref{sec_stardisc} enables one
to calculate the net torque ($\tacc + \tmag$) on the star for any
accreting system with known $M_*$, $R_*$, $\mu$, $\mdotacc$, and
$\Omega_*$ (one must also adopt a value for $\gammacrit$ and $\beta$).
The system is stable, in that a positive torque spins the star up, and
a faster spin reduces the total torque.  Conversely, a negative torque
spins down the star, and the torque increases (becoming less negative)
for slower spin.  Therefore, in a system where the other parameters
are relatively constant, the spin rate of the star naturally adjusts
to an equilibrium state in which $\tacc + \tmag = 0$, known as the
`disc-locked' state (e.g., K91; \citealp{cameroncampbell93, shuea94};
AC96).  Since the only torques that spin down the star originate along
field lines that connect to the disc outside $\rco$, systems in their
equilibrium state must be in state 2 ($\rt > \rin$, see Fig.\
\ref{fig_states}).  In this section, we show that both $\rt$ and
$\Omega_*$ in the disc-locked state are significantly affected by the
field topology and thus have a strong dependence on the magnetic
diffusion parameter $\beta$.

     \subsection{Truncation radius in the disc-locked state} \label{sub_rteq}

The disc-locked state is defined by the condition, $\tmag = -\tacc$.
Thus, by combining equations \ref{eqn_tacc} and \ref{eqn_tmag}, and
using equations \ref{eqn_rco} and \ref{eqn_rt} to eliminate $f$, this
condition can be rearranged to be
\begin{eqnarray}
\label{eqn_rteq}
{{{K(\beta \gammacrit) - 
  \left({\rco / \rt}\right)^{3/2}_{\rm eq}}} \over
  \left({\rco / \rt}\right)^{{3/2}}_{\rm eq}
  \left[{1 - \left({\rco / \rt}\right)^{3/2}_{\rm eq}}\right]} = 7,
\end{eqnarray}
where
\begin{eqnarray}
\label{eqn_kbg}
K(\beta \gammacrit) \equiv 2 (1 + \beta \gammacrit)^{-1} - 
 (1 + \beta \gammacrit)^{-2},
\end{eqnarray}
and the subscript `eq' refers to the value in the disc-locked state.
In deriving equation \ref{eqn_rteq}, we have ignored the term
proportional to $k^2 f$ in equation \ref{eqn_tacc}, as justified in
the discussion following that equation.  The function $K(\beta
\gammacrit)$ characterizes the topology of the field in a sense that,
when $\beta \gammacrit$ varies between 0 and $\infty$, $K$ varies
between 1 (completely open field) and 0 (completely closed).  Equation
\ref{eqn_rteq} has exactly one solution such that $(\rt/\rco)_{\rm eq}
< 1$ (which is the only physical solution) for any given $\beta
\gammacrit > 0$.

The location of $\rt$ for accreting systems is, in principle, an
observable parameter.  For example, \citet{kenyon3ea96} used a
magnetic accretion model to predict infrared excesses in CTTS's and
then to determine the value of $\rt/\rco$ for a sample of stars in the
Taurus-Auriga molecular cloud.  The value of $(\rt/\rco)_{\rm eq}$
predicted by equation \ref{eqn_rteq} represents the value for a system
that is disc locked.

\begin{figure}
\centerline{\includegraphics[width=22pc]{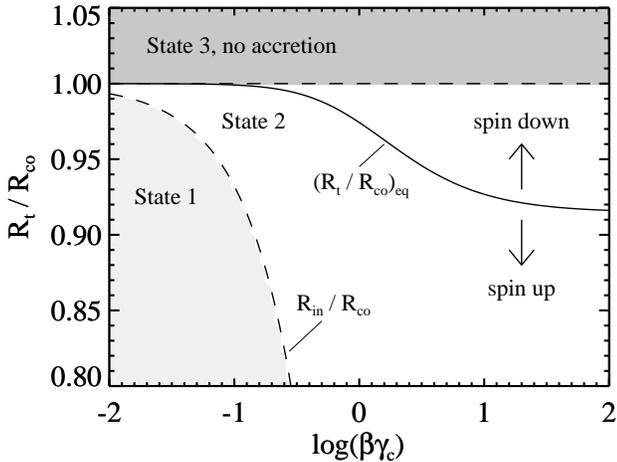}}

\caption{Location of $\rt/\rco$ in the disc-locked state (solid line;
eq.\ \ref{eqn_rteq}), as function of $\log(\beta \gammacrit)$.  The
dashed lines represent $\rin/\rco$ and the location where $\rt/\rco =
1$.  Any system with $\rt > \rco$ is in state 3 (dark grey shaded
region), while any system with $\rt < \rin$ is in state 1 (light
grey).  Furthermore, if $\rt/\rco$ is anywhere below (above) the solid
line, the net torque from the disc will spin the star up (down).
\label{fig_rteq_rcor}}

\end{figure}

The solid line in Figure \ref{fig_rteq_rcor} shows $(\rt/\rco)_{\rm
eq}$ as a function of $\log(\beta \gammacrit)$.  When a closed field
topology is assumed ($\gammacrit \rightarrow \infty$),
$(\rt/\rco)_{\rm eq} \approx 0.915$.  However, for the more reasonable
value of $\gammacrit = 1$, $(\rt/\rco)_{\rm eq}$ increases
(approaching unity) as the magnetic coupling becomes stronger (smaller
$\beta$).  This confirms the conclusion of \citet[][and see
\citealp{cameroncampbell93, yi94, yi95}]{wang95} that any disc-locked
system will have $(\rt/\rco)_{\rm eq} \ga 0.9$, and we find that the
effect of a more open field topology is to significantly increase this
value.

The Figure also indicates the three possible states of the system
(discussed further in \S \ref{sub_3states}), determined by the
location of $\rt$ relative to $\rin$ and $\rco$ (dashed lines).  A
system that is disc-locked is always in state 2.  If a system is
observed with $\rt/\rco$ larger than the solid line in the Figure, the
star should be spinning down.  Conversely, if $\rt/\rco$ is smaller
than the solid line in the Figure, the net torque from accretion and
from field lines connecting the star and disc will act to spin the
star up.  It is interesting that \citet{kenyon3ea96} found typical
values of $\rt/\rco$ in the range of 0.6 to 0.8 for the stars in their
sample.  If true, these stars cannot be in spin equilibrium, unless
they feel significant spin-down torques {\it other than} those from
field lines connecting them to their discs.  Furthermore, if $\beta
\gammacrit < 0.3$ is appropriate, the stars in their sample should
exist in state 1.

     \subsection{Stellar spin rate in equilibrium} \label{sub_spineq}

Now that we can calculate $(\rt/\rco)_{\rm eq}$ via equation
\ref{eqn_rteq}, we rewrite equation \ref{eqn_rt} to find the
fractional spin rate of the star in equilibrium
\begin{eqnarray}
\label{eqn_feq}
f_{\rm eq} = C(\beta, \gammacrit) (2/\psi)^{3/7},
\end{eqnarray}
where
\begin{eqnarray}
\label{eqn_cbg}
C(\beta, \gammacrit) \equiv
  \left\{{{2\over \beta}
  \left({\rco \over \rt}\right)^{2}_{\rm eq}
  \left[{\left({\rco \over \rt}\right)^{3/2}_{\rm eq} - 1}\right]
  }\right\}^{-3/7}.
\end{eqnarray}
Since $(\rco / \rt)_{\rm eq}$ depends only on $\beta \gammacrit$ (via
eqs.\ \ref{eqn_rteq} and \ref{eqn_kbg}), the dimensionless function
$C(\beta, \gammacrit)$ depends only on $\beta$ and $\gammacrit$.  We
can also combine equations \ref{eqn_f}, \ref{eqn_psi}, and
\ref{eqn_feq} to find the equilibrium angular spin rate of the star
\begin{eqnarray}
\label{eqn_spineq}
\omegaeq = C(\beta, \gammacrit)
  \mdotacc^{3/7} \left({G M_*}\right)^{5/7} \mu^{-{6/7}}.
\end{eqnarray}
This equation has the same dependence of $\omegaeq$ on $\mdotacc$,
$M_*$, and $\mu$ as equation 3 of K91, and as in the theory of
\citet[][and \citealp{ostrikershu95}]{shuea94}.  The only difference
is the value of the factor $C$ used in the various theories.  K91 used
$C \approx 1.15$, and \citet{ostrikershu95}\footnote{ While it is
interesting that our equation \ref{eqn_spineq} resembles the
formulation of \citet{ostrikershu95}, their assumed magnetic field
geometry is different than ours, so the comparison of $C$ values
should not be taken too seriously.}  found $C \approx 1.13$.  However,
our formulation of the problem allows us to determine the effect of
the field topology on the equilibrium spin rate, via the function
$C(\beta, \gammacrit)$, for arbitrary values of the diffusion
parameter $\beta$.

\begin{figure}
\centerline{\includegraphics[width=22pc]{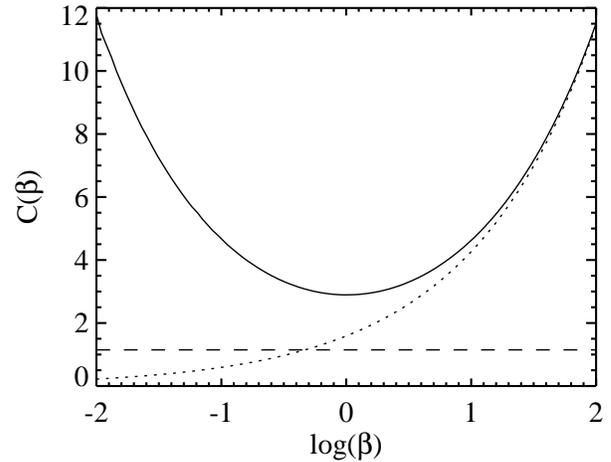}}

\caption{Spin rate factor $C(\beta, \gammacrit)$ (eq.\ \ref{eqn_cbg})
as a function of $\log(\beta)$.  The dotted line is $C(\beta)$ for
$\gammacrit \rightarrow \infty$ and represents the assumption of a
completely closed magnetic topology.  The solid line shows the more
realistic case with a partially open topology ($\gammacrit = 1$).  The
dashed lines represents $C \approx 1.15$ from K91.  The Figure is from
\citet{mattpudritz04a}.
\label{fig_cbg}}

\end{figure}

Figure \ref{fig_cbg} reveals the dependence of $C(\beta, \gammacrit)$
on $\beta$ for the two values of $\gammacrit$ we have considered
throughout.  For $\gammacrit \rightarrow \infty$, $C(\beta) \approx
1.59 \beta^{3/7}$, which is represented by the dotted line in the
Figure.  The solid line shows $C(\beta, \gammacrit)$ for the more
realistic value of $\gammacrit = 1$ and illustrates the effect of a
reduced magnetic connection to the disc.  For comparison, the dashed
line shows the spin rate factor used by K91, which also roughly
represents the typical factors of order unity used in most
star-disc interaction models.

The dotted line in Figure \ref{fig_cbg} represents the assumption that
the magnetic field everywhere connects to the disc, regardless of the
field twist.  In that case, the magnetic torque increases with with
decreasing $\beta$, since the field then becomes highly twisted, and
so the prediction is that $\omegaeq \propto \beta^{3/7}$.  However,
when one considers that dipole field lines will become open when
largely twisted, the torque has a maximum value for $\beta = 1$ and
decreases for any other $\beta$ (see \S \ref{sec_open}).
Correspondingly, the solid line in Figure \ref{fig_cbg} has a minimum
value for $\beta = 1$.  This minimum value represents the `best
case' for disc locking, and even at this location, $C$ is a factor of
1.8 larger than the value for $\gammacrit \rightarrow \infty$ and 2.5
times larger than used by K91.  For the more likely value of $\beta =
10^{-2}$, the predicted equilibrium spin rate of the star is more than
an order of magnitude faster than prediced by any other model.  Note
that the spin rates of the system plotted in Figures
\ref{fig_dtorques} and \ref{fig_dtorques2} were chosen to be in
equilibrium, and a comparison between the two Figures shows the effect
of field topology for the $\beta = 1$ `best case.'

\begin{figure*}
\centerline{\includegraphics[width=36pc]{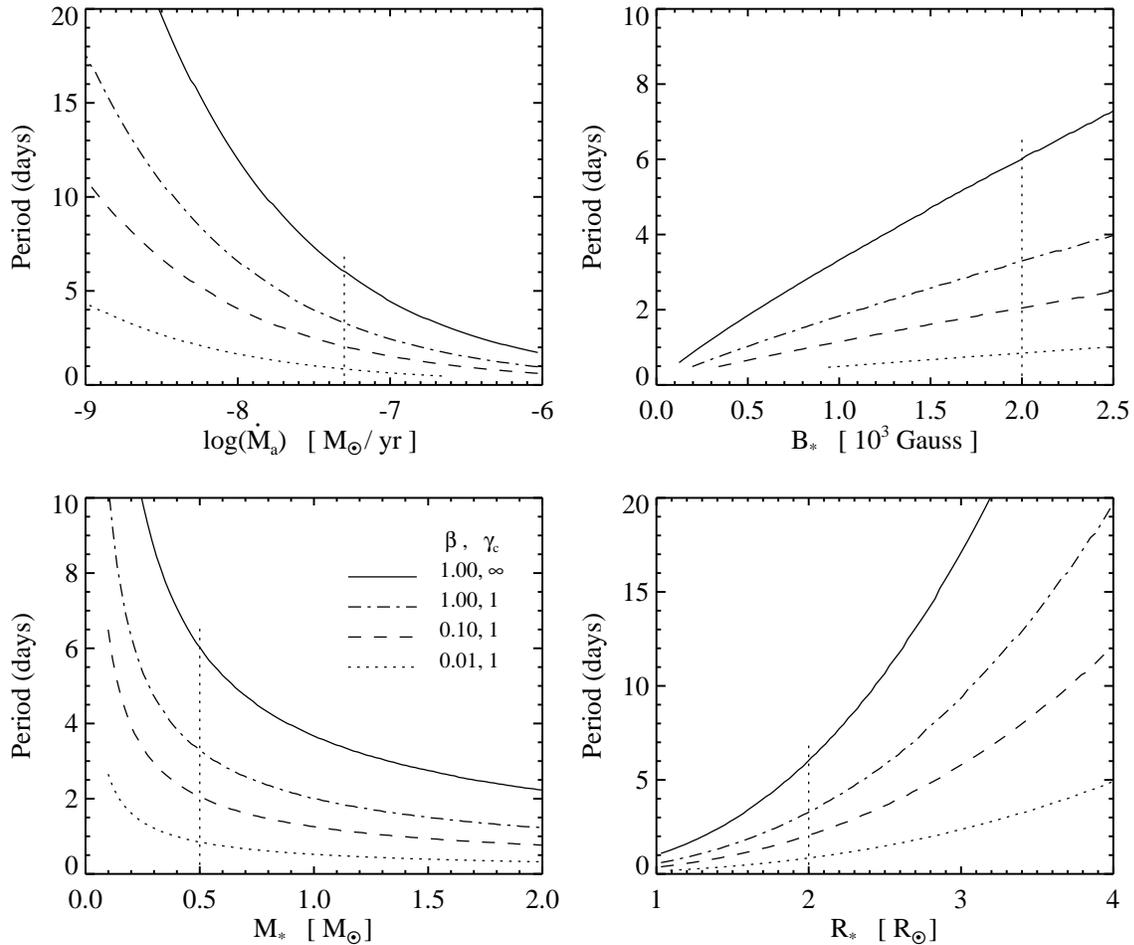}}

\caption{Equilibrium spin period as a function of observable
parameters $\mdotacc$ (upper left), $B_*$ (upper right), $M_*$ (lower
left), and $R_*$ (lower right).  In each panel, all parameters are
held fixed at the fiducial value except for that which is plotted
along the abscissa.  The vertical dotted line in each panel shows the
fiducial value that is held fixed in the other panels.  The solid line
in each panel represents the assumption that the magnetic field
topology is completely closed ($\gammacrit \rightarrow \infty$) and
$\beta = 1$.  All of the broken lines represent a partially open
topology ($\gammacrit = 1$) for $\beta = 1$ (dash-dot lines), 0.1
(dashed lines), and the more likely value of 0.01 (dotted lines).
\label{fig_fequil}}

\end{figure*}

\citet{mattpudritz04a} applied the analysis presented in this section
to the CTTS BP Tau, which is one of the few stars for which all of the
relevant system parameters are known (or well-constrained).  They
argued that the existence of slowly rotating, accreting stars, such as
BP Tau, cannot be explained by a disc-locking scenario.  To further
illustrate the effect of field topology on the predictions of
disc-locking, and to apply the analysis to all CTTS's, we have plotted
Figure \ref{fig_fequil}.  The Figure shows the predicted spin period
for a wide range of observable parameters.  The spin period is given
by $2 \pi / \omegaeq$, and we have used the relationship $\mu = B_*
R_*^3$.

The solid lines are for $\gammacrit \rightarrow \infty$ and $\beta =
1$, and so they represent the `standard prediction' by previous
models.  The broken lines take into account that some of the field
should be open ($\gammacrit = 1$) for the three different values of
$\beta = 1$, 0.1, and 0.01.  Note that $\beta = 0.1$ predicts the same
period as for $\beta = 10$ (due to the approximate symmetry of the
solid line in Fig.\ \ref{fig_cbg}), and $\beta = 0.01$ corresponds
with $\beta = 100$.  It is evident that, even in the `best case'
($\beta = 1$) for the disc-locking scenario, the effect of a more open
topology is to reduce the equilibrium spin period by a factor of two,
compared to the closed field assumption.  Given the uncertainties in
some of the observed parameters, it may not yet be possible to
constrain the predicted period to within a factor of two.  In
particular, a difference of a factor of two in the predicted spin
period could result from an error of a factor of 5.0, 2.2, 2.6, or 1.3
in the observed parameters $\mdotacc$, $B_*$, $M_*$, or $R_*$,
respectively.  However, for the more likely value of $\beta = 10^{-2}$
(dotted lines in Fig.\ \ref{fig_fequil}), the predicted spin period is
an order of magnitude lower than the `standard prediction,' which
cannot be reconciled by observational errors.

     \subsection{Time to reach equilibrium} \label{sub_time}

It is important to determine how quickly the star will spin up or down
to reach the equilibrium state.  Rather than fully solving the
time-dependent problem, which should also include the spin-up due to
the contraction of the protostar \citep[e.g.,][]{yi94}, one typically
estimates a characteristic spin-down time using the angular momentum
of the star, $L_*$, divided by the net torque on the star.  To be more
precise, and for arbitrary spin rates, one should replace $L_*$ with
the difference between the current $L_*$ and the value of $L_*$ for
the equilibrium spin rate.  Assuming solid body rotation of the star,
the characteristic time to reach spin equilibrium is then
\begin{eqnarray}
\label{eqn_time}
t_{\rm spin} = 
  M_* k^2 R_*^2 
  \left({{\omegaeq - \Omega_*} \over {\tacc + \tmag}}\right),
\end{eqnarray}
which corresponds to a spin-up (down) time for a star currently
spinning slower (faster) than $\omegaeq$.  If $t_{\rm spin}$ is long
compared to the time-scale for other system parameters to change (e.g.,
compared to the lifetime of the disc), the star is unlikely to ever be
in a spin equilibrium state.

\begin{figure}
\centerline{\includegraphics[width=22pc]{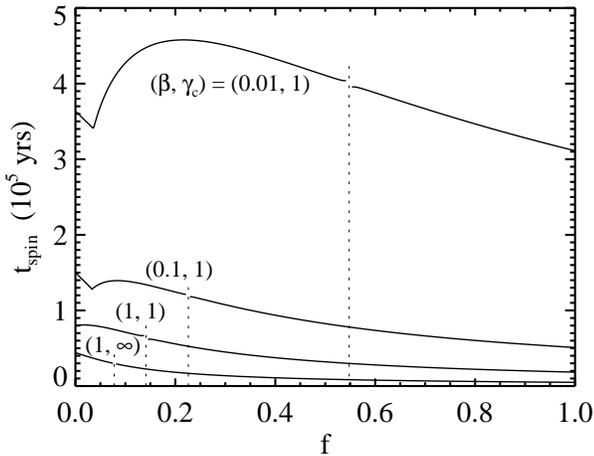}}

\caption{Time to reach equilibrium, $t_{\rm spin}$ (eq.\
\ref{eqn_time}), as a function of the spin rate fraction $f$, for the
fiducial CTTS system.  The four solid lines represent $(\beta,
\gammacrit) = (1, \infty)$, $(1, 1)$, $(0.1, 1)$, and $(0.01, 1)$, as
indicated.  The vertical dotted lines indicate the equilibrium spin
rate, for each case.
\label{fig_sdtimes}}

\end{figure}


Figure \ref{fig_sdtimes} shows $t_{\rm spin}$ for the adopted fiducial
parameters, as a function of the spin rate fraction $f$ and for
different values of $\beta$ and $\gammacrit$.  It is evident from the
Figure that $t_{\rm spin}$ generally decreases for increasing $f$,
since faster spin means $\rco$ is closer to the star so the magnetic
torque will be stronger (whether it spins the star up or down).  There
is an exception to this when the star spins much slower than $f_{\rm
eq}$.  For example, for the case with $(\beta, \gammacrit)$ = (0.01,
1), when $f \la 0.2$, $t_{\rm spin}$ decreases with decreasing $f$.
This can be understood, since then $\rt/\rco$ decreases rapidly with
decreasing $f$ (see Fig.\ \ref{fig_rtrstar}), leading to much stronger
magnetic spin-up torques.  Note also that the cases with $(\beta,
\gammacrit)$ = (0.01, 1) and (0.1, 1) are in state 1 for $f \la 0.036$
and 0.033, respectively, in which $\tmag = 0$.

For the `standard' prediction with $(\beta, \gammacrit)$ = (1,
$\infty$), $t_{\rm spin}$ is always less than $5 \times 10^4$ years,
which is much shorter than the expected disc lifetime of more than
$10^6$ years \citep{muzerolle3ea00}.  However, when one considers the
effect of a more open field topology ($\gammacrit = 1$), the magnetic
torques is reduced, and so $t_{\rm spin}$ is longer and increases when
$\beta$ decreases.  For the likely case with $(\beta, \gammacrit)$ =
(0.01, 1), $t_{\rm spin} \sim 4 \times 10^5$ years.  This is still
relatively short compared to the expected disc lifetime.  Therefore,
we agree with previous authors (e.g., K91; AC96) that systems such as
those considered above should exist near their equilibrium spin states
throughout most of their accretion lifetimes, but only if $\beta \ga
0.01$.  However, the effect of a more open field topology is that the
equilibrium spin rate is much faster than previously predicted.

We note in passing that the characteristic time $t_{\rm spin}$ we have
calculated here is significantly shorter than recently calculated by
\citet{hartmann02}.  Hartmann's estimate assumes that the upper limit
to the spin-down torque on the star is equal to $\mdotacc (G M_*
\rout)^{1/2}$.  However, as discussed in section \ref{sub_torques},
this is the torque necessary to provide a steady accretion rate of
$\mdotacc$ through the radius $\rout$.  In order for the disc to exert
a spin-down torque on the star (via the magnetic connection), it must
provide a torque {\it in addition to} $\mdotacc (G M_* \rout)^{1/2}$,
requiring the disc to have a different structure than in the absence
of a stellar field \citep[see][]{rappaport3ea04}.  It is not yet clear
to what extent the disc can be restructured (before accretion will
cease), and so Hartmann's estimate does not represent a true limit.

\section{Discussion and conclusions} \label{sec_discussion}

We extended the standard picture of the interaction of magnetized star
with a steady-state accretion disc.  Our more comprehensive
formulation of this problem allows us to determine the location of the
disc truncation radius $\rt$ and calculate the torque on the star for
a system with arbitrary values of $\mdotacc$, $\Omega_*$, $M_*$,
$R_*$, $B_*$, and $\beta$, which parametrizes the coupling of the
magnetic field to the disc.  We consider only two sources for the
torques: (a) torque from the angular momentum deposited by accretion
of disc material from $\rt$, and (b) torques exerted by field lines
connecting the star to the disc over the region from $\rt$ to $\rout$.

Our model resembles several previous studies (e.g., AC96), except that
we have now determined the dependence of the torques on the magnetic
coupling to the disc.  Specifically, the differential rotation between
the star and disc results in a largely open topology (e.g., UKL), so
the size of the region of the disc that is magnetically connected to
the star is smaller (i.e., $\rout$ is smaller).  Thus, when one
considers this effect, the magnetic spin-down torque on the star is
less than if one assumes the field remains everywhere closed.  The
strongest spin-down torque occurs for intermediate magnetic field
coupling ($\beta = 1$), in which case the spin-down torque is a factor
of four less than for the closed field assumption.  For strong
magnetic coupling ($\beta \ll 1$), as expected near the inner edge of
an accretion disc, $\rout$ is very close to $\rco$, and the spin-down
torque then is proportional to $\beta$.  For the likely value of
$\beta = 0.01$, the spin-down torque is 100 times less than for the
closed field assumption!  Furthermore, the possibility that field
lines may open inside $\rco$ characterizes a new mode (state 1) in the
system, in which the star feels no spin-down torques from field lines
connected to the disc.  Three possible system states are summarized in
section \ref{sub_3states}.

We also considered the disc-locked state of the system, in which the
net torque on the star is zero.  A more open field topology leads to
an equilibrium state that has a higher stellar spin rate.  The time
for a given system to reach spin equilibrium is also longer when the
field is more open.  These results apply to any system in which
accretion occurs onto a magnetized central object.  In the general
case, not all of the system parameters are observationally known.
Thus, one often assumes that a particular system is disc-locked, and
then `tunes' the unknown system parameter(s) to satisfy equation
\ref{eqn_spineq}.  We found that, equation \ref{eqn_spineq} contains
the function $C(\beta, \gammacrit)$, which is plotted in Figure
\ref{fig_cbg} (solid line), as a function of $\beta$.  It is evident,
that when the coupling of the field to the disc is strong ($\beta \ll
1$) or weak ($\beta \gg 1$), the `tunable' system parameters will be
significantly different than for the usual assumption that $C(\beta,
\gammacrit)$ is a constant near unity.  In particular, systems that
are thought to be disc-locked will require a larger $\mu$, larger
$\omegaeq$, smaller $M_*$, or smaller $\mdotacc$ than calculated using
previous formulations of equation \ref{eqn_spineq}.  In the next
section, we discuss the implications of these results, and recent
results from the literature, for disc-locking in CTTS's.

     \subsection{Problems with disc locking for CTTS's} \label{sub_problems}

Observational support for disc locking in CTTS's is still
controversial \citep[in particualar, see][]{stassunea99, herbstea00,
stassunea01, herbstea02}, so we have taken another look at the problem
from a theoretical standpoint.  We find that spin-down torques on a
CTTS are significantly reduced for strong coupling of the stellar
field to the disc (small $\beta$), resulting in equilibrium spin
periods as low as a few days or less, for a wide range of system
parameters (see Fig.\ \ref{fig_fequil}).  Small values of $\beta$ are
likely for CTTS's (see \S \ref{sec_twist}), although $\beta$ is a
highly uncertain parameter to calculate from first principles and may
even vary from system to system.  Also, as discussed in section
\ref{sub_tmaxtwist}, there may exist special circumstances that allow
the field to remain connected, but whether these circumstances can
exist in CTTS systems remains to be shown.  Therefore, while our
analysis of the problem (in \S \ref{sec_equilib}) does not completely
rule out the possibility for disc locking in all systems (particularly
for fast rotators), it significantly reduces the likelyhood that
disc-locking can explain the existence of accreting stars spinning at
$\sim10$\% \citep[e.g.,][]{bouvierea93} of break-up speed.  There are
several other issues in recent literature that, when combined with our
analysis, cast additional doubt on the applicability of disc-locking
to the slow rotators \footnote{ Due to the the unique field geometry
of the X-wind, in which all the stellar field lines are squeezed into
the inner edge of the disc \citep{shuea94}, that model avoids the
problem of field opening due to differential twisting, as considered
in this paper.  However, the issues discussed after the first
paragraph in this section apply to all disc-locking models, including
the X-wind.}.

The most notable observational challenge to the disc-locking model is
the apparent lack of strong dipole fields on CTTS's \citep[first
suggested by][]{safier98}.  It is generally accepted that CTTS's have
field strengths of a few kiloguass.  This was predicted by K91, and
subsequent observations \citep{basri3ea92, guentherea99,
johnskrull3ea99, johnskrullvalenti00, johnskrullea01} have indeed
found a {\it mean} field on the surface of the central stars of
typically 2 kG.  Thus far in our analysis, we have adopted a field
strength of 2 kG to represent a typical CTTS system and guide our
discussion.  However, stringent measurements of the mean {\it line of
sight} field have been carried out for three CTTS's, BP Tau
\citep{johnskrullea99}, TW Hya \citep{johnskrullvalenti01}, and T Tau
\citep{smirnovea03, smirnovea04}, and all measurements give an upper
limit of roughly 200 G for the strength of the dipole component of the
stellar magnetic field.  The measured mean fields of 2 kG thus
represent a field that is disordered or characterized by multipoles of
higher order than a dipole \citep{johnskrullea99}.  Such high order
fields, even if very strong on the stellar surface, decrease in
strength too quickly with increasing radius to exert significant
spin-down torques, especially for slow rotators in which $\rco$ is at
several stellar radii.  Furthermore, a 200 G dipole field cannot exert
a significant spin-down torque on a CTTS, even if the field connects
to the disc everywhere outside $\rco$.  This is evident, for example,
in the upper right panel of Figure \ref{fig_fequil}, which indicates
that the equilibrium spin period for a star with such a field is less
than one day.  For the cases with $\gammacrit = 1$ and $\beta \la
0.1$, there is no equilibrium possible, since the magnetic spin-down
torques are not strong enough to counteract the angular momentum added
by accretion, even for maximal stellar spin ($f = 1$).  Also, the time
to reach equilibrium (for cases in which equilibrium is possible; as
discussed in \S \ref{sub_time}) increases by an order of magnitude
when $B_*$ = 200 G, compared to 2 kG.

Second is the issue pointed out by \citet{wang95} and discussed in
section \ref{sub_rteq} that stars in their disc-locked state must have
$\rt/\rco \ga 0.9$, for a wide range of possible assumptions in the
model.  Interestingly, \citet{kenyon3ea96} concluded that, for their
CTTS sample, the typical value of $\rt/\rco$ was in the range of 0.6
to 0.8, well below the disc-locked value.  If true, these stars cannot
be disc-locked, since the sum of the accretion torque and the torque
carried by field lines connected to the disc will be positive
(spinning the stars up; see \S \ref{sub_tmag} and \ref{sub_rteq}).
Thus, these stars can only be in spin equilibrium if they feel
significant spin-down torques {\it other than} those from field lines
connecting them to their discs.  This conclusion does not depend on
whether or not the stellar field can open, since the calculation of
$\rt$ in equilibrium also does not.

Finally, CTTS's may drive stellar winds \citep{safier98}, and outflows
from the disc are known to explain several aspects of observed
protostellar outflows \citep[e.g.][]{koniglpudritz00}.  Ionized winds
escape from regions with open magnetic field lines, or they can
themselves open the field \citep[e.g., as in the solar
wind;][]{parker58}, disconnecting the star and disc.  \citet{safier98}
concluded that CTTS's winds should open all stellar field lines beyond
roughly 3 $R_*$.  Furthermore, a recent measurement of rotation in the
jet from the CTTS DG Tau \citep[][and see
\citealp{testiea02}]{bacciottiea02} suggests that the low velocity
component ($\sim 70$ km s$^{-1}$) originates in the disc from as close
as 0.3 AU from the star \citep[][]{andersonea03}.  This is likely an
upper limit \citep{presentiea04}, and the more tightly collimated,
high velocity component \citep[$\sim 220$ km s$^{-1}$;][]{pyoea03}
must then originate from well within this radius in the disc
\citep{andersonea03}.  Theoretical disc-wind models
\citep[e.g.,][]{koniglpudritz00} predict jet speeds of the order of
the Keplerian velocity from where the wind is launched, and observed
protostellar jets typically travel with speeds of a few hundred km
s$^{-1}$ \citep[e.g.,][]{reipurthbally01}.  Considering a star with
$M_* = 0.5 M_\odot$ and $R_* = 2 R_\odot$, and assuming $v_{\rm k} =
100$ km s$^{-1}$ at the launch point, this requires the disc winds to
originate from near $GM_*/v_{\rm k}^2 \approx 4.8 R_*$.  These ionized
disc winds prevent the star from being magnetically connected to the
disc beyond the innermost location of the wind launching point
(effectively giving an upper limit to $\rout$), and these winds carry
angular momentum from the disc, not from the star.  As calculated in
this paper, a reduced size of the connected region leads to a reduced
spin-down torque on the star, regardless of the cause of the field
opening.  Protostellar outflows thus provide an independent and
stringent constraint on disc-locking models.

We conclude that spin-down torques exerted by field lines connecting
the star to the disc outside $\rco$ are likely to be much weaker than
usually assumed.  Therefore, the existence of slowly rotating ($f \la
0.1$, or perhaps higher) CTTS's probably cannot be explained by a
disc-locking scenario.  Either these stars are all in the process of
spinning up, or the stars feel torques other than those related to a
magnetic connnection to the accretion disc.  Given that the typical
spin-up times for these systems are short (see \S \ref{sub_time}), the
latter possibility appears the most likely.

     \subsection{Three states of the system} \label{sub_3states}

We identified three possible configurations of the system, determined
by the location of $\rt$, relative to the two key radii $\rin$ (eq.\
\ref{eqn_rin}) and $\rco$.  There are two accreting configurations,
which we call states 1 and 2, and one non-accreting configuration,
state 3.  Here, we summarize the conditions that determine and
characterize each state.


Figure \ref{fig_states} illustrates the basic magnetic configuration
of the three states, which may even represent an evolutionary sequence
for a system with (e.g.) an evolving $\mdotacc$ (see \S
\ref{sub_states}).  Figure \ref{fig_rteq_rcor} shows the location of
$\rin/\rco$ (curved, dashed line) for various values of $\beta
\gammacrit$, and the horizontal dashed line represents the location of
$\rco$.  For a system with a given value of $\beta \gammacrit$ and for
which $\rt$ is determined, the Figure indicates which state the system
will be in and whether the star should be spinning up or down.

          \subsubsection{State 1: $\rt < \rin$ (or $\rout^\prime < \rco$)} \label{sub_state1}

This state is a direct consequence of the opening of field lines via
the differential rotation between the star and disc.  When $\rt$ is
sufficiently less than $\rco$ (i.e., when $\rt < \rin$), the magnetic
field becomes highly twisted there, opening the field inside $\rco$
and resulting in a highly open field topology.  Note that state 1, in
this context, is only possible for $\beta \gammacrit < 1$, since
otherwise $\rin$ is undefined (see Fig.\ \ref{fig_rteq_rcor} and
discussion in \S \ref{sub_rinrout}).

As illustrated in the top panel of Figure \ref{fig_states} and
discussed in section \ref{sub_rt}, the stellar field in state 1
connects only to a very small, innermost region of the disc near
$\rt$, and all exterior field lines are open.  The size of the small
connected region is likely to be determined by dissipative processes
within the disc, but we do not attempt to calculate this here.  The
location of $\rt$ is determined by equation \ref{eqn_rt2} (and see
Fig.\ \ref{fig_rtrstar}), and accretion onto the star occurs from
there.

A star in this state will always feel a net positive torque from the
disc, since no field connects outside $\rco$.  Specifically, the star
is spun up by the accretion torque ($\tacc$; eq.\ \ref{eqn_tacc}), and
equation \ref{eqn_tmag} for the magnetic torque is not applicable.  In
fact, since $\rt$ increases with stellar field strength, and since
$\tacc \propto \rt^{1/2}$, the presence of a stellar field actually
{\it increases} the spin up torque on the star, relative to
non-magnetic accretion.  In any case, a system in state 1 cannot be in
spin equilibrium, unless it receives torques other than those
considered in this paper.

Is state 1 a likely, or even common, configuration for accreting
systems?  Figure \ref{fig_rteq_rcor} indicates that, when the magnetic
coupling to the disc is strong, the range of possible values of $\rt$
for which a system can be in state 2 is significantly reduced.  For
example, if $\beta \gammacrit = 0.01$, any system with $\rt < 0.993
\rco$ should be in state 1.  The specific conditions under which any
given system will be in state 1 is given by equation \ref{eqn_trans}.
For illustrative purposes, we can solve this equation (and using eq.\
\ref{eqn_psi}) for the mass accretion rate.  Assuming $\gammacrit =
1$, $\beta = 0.01$, and $f = 0.1$, and using the fiducial CTTS values
(see discussion above \S \ref{sec_twist}), one finds that a system
should be in state 1 if
\begin{eqnarray}
\label{eqn_mtrans}
\mdotacc > 5.4 \times 10^{-7} 
  \left({\gammacrit \over 1}\right)
  \left({1 - \beta \gammacrit \over 0.99}\right)^{-7/3}
  \left({M_* \over 0.5 \; M_\odot}\right)^{-1/2} \nonumber \\
  \left({B_* \over 2 \; {\rm kG}}\right)^{2}
  \left({R_* \over 2 \; R_\odot}\right)^{5/2}
  \left({f \over 0.1}\right)^{7/3}
  M_\odot \; {\rm yr}^{-1}.
\end{eqnarray}
This threshold value of $\mdotacc$ is an order of magnitude larger
than the fiducual value,
suggesting that CTTS slow rotators will most commonly exist in state
2.

However, equation \ref{eqn_mtrans} assumes a magnetic field strength
of 2 kG, and as discussed in section \ref{sub_problems}, the stars
likely have surface dipole field strengths of less than 200 G. This
consideration decreases the threshold value of $\mdotacc$ by at least
two orders of magnitude, suggesting that typical CTTS systems may
exist in state 1.  We can also look at this from the standpoint of
stellar spin, using equation \ref{eqn_trans}, which indicates that a
system with the adopted fiducial parameters (but with $B_* = 200$ G)
will be in state 1 if it spins more slowly than 26\% of break-up speed.
(As discussed in section \ref{sub_rt}, this also corresponds an upper
limit of $\rt \la 2.4 R_*$.)  Thus, if the dipole fields are indeed
weak, it is more likely that slow rotators, and even some fast
rotators, will be in state 1.  Furthermore, the conclusion of
\citet[][]{kenyon3ea96}, that $\rt/\rco$ typically ranges from 0.6 to
0.8, suggests that CTTS's will be in state 1, as long as $\beta
\gammacrit < 0.3$ (see \S \ref{sub_rteq}).

We have thus far considered the opening of field lines via the
differential rotation, so the existence of state 1 requires that
$\beta \gammacrit$ is significantly less than unity.  Given the large
uncertainty in the value of $\beta$ in real systems, it is still not
clear whether or not state 1 should be common.  From the standpoint of
torques on the star, the most important feature of state 1 is that the
stellar field never connects to the disc outside $\rco$.  Thus, for
the following discussion, we will generalize the definition of state 1
to include any magnetic configuration in which the star does not
connect outside $\rco$.

State 1 is characterized by a large amount of open stellar field, so
it is natural to consider the effects of a stellar wind in the
magnetically open region.  As discussed in section \ref{sub_problems},
a wind can even be responsible for opening the field (which does not
depend on our parameters $\beta$ and $\gammacrit$).  Thus, if a wind
(or any other process) keeps the stellar field open beyond some radius
$\rout^\prime$, and if $\rout^\prime < \rco$, the system will be in
state 1.  For example, \citet{safier98} concluded that stellar winds
from CTTS's could result in $\rout^\prime \la 3 R_*$.  If true, this
means that any system rotating more slowly than 19\% of break-up speed
will have $\rout^\prime < \rco$ (eq.\ \ref{eqn_f}) and be in state 1.

There is empirical evidence for systems in state 1 from some numerical
simulations of the star-disc interaction, which usually represent
systems with $\beta \ll 1$.  In the simulations of
\citet{goodsonwinglee99} and \citet{vonrekowskibrandenburg04}, as an
example, after the initial state, the stellar field never connects to
the disc outside $\rco$, even immediately following reconnection
events.  These authors report that the only significant spin-down
torques on the star come from the open field regions (though a stellar
wind was not properly included), rather than along field lines
connecting the stars to their discs \citep[but also
see][]{romanovaea02}.

It seems that state 1 is a likely configuration for accreting stars,
particularly among slow rotators.  Since, in this state, the net
torque from the interaction with the accretion disc only acts to spin
up the star, stars with long-lived accretion phases must somehow rid
themselves of this excess angular momentum.  Stellar winds can exert
spin-down torques on the star, and if these torques are significant
\citep[e.g.,][]{toutpringle92}, the equilibrium spin rate may be
simply determined by a balance between this torque and $\tacc$.  In
this situation, state 1 could actually represent the expected
configuration for accreting systems in spin equilibrium.

         \subsubsection{State 2: $\rin < \rt < \rco$} \label{sub_state2}

In this state, the stellar field connects to a finite region of the
disc between $\rt$ and $\rout$, as illustrated in the middle panel of
Figure \ref{fig_states}.  This represents the typical configuration in
many star-disc interaction models, except that the determination of
$\rout$ varies between models.  The location of $\rt$ is determined by
equation \ref{eqn_rt} (and see Fig.\ \ref{fig_rtrstar}), and accretion
onto the star occurs from there.

The star is spun up by the accretion torque (eq.\ \ref{eqn_tacc}) and
magnetic torques (eq.\ \ref{eqn_tmag}) from field lines connected to
the region of the disc between $\rt$ and $\rco$ and spun down by
magnetic torques from field lines connected between $\rco$ and
$\rout$.  Therefore, a system can exist in an equilibrium, disc-locked
state, in state 2, in which the net torque on the star is zero, and
the spin rate of the star then correlates with accretion parameters
(see \S \ref{sec_equilib}).

When one considers that the differential rotation determines $\rout$
(via eq.\ \ref{eqn_rout}), both $\omegaeq$ and $(\rt/\rco)_{\rm eq}$
are larger for smaller $\beta$.  Also, as shown in Figure
\ref{fig_rteq_rcor}, the range of (non-equilibrium) values of $\rt$
that exist in state 2 becomes narrower as $\beta$ decreases.  For the
strong coupling case of $\beta \gammacrit = 0.01$, a system can only
be in state 2 if $0.993 < \rt/\rco < 1.0$.  So for strong coupling, it
is unlikely that a given system will exist in state 2, unless it is
very near its disc-locked state, which then requires a fast stellar
spin.

Another intriguing effect of a largely open field topology is that, in
order for a system to be disc-locked, the differential magnetic torque
in the disc $d \tmag$ must be stronger (compared to the completely
closed assumption) in order to make up for the decreased size of the
connected region (compare Figs.\ \ref{fig_dtorques} and
\ref{fig_dtorques2}).  When the connected region is very small (i.e.,
for small $\beta$), $d \tmag$ is very large, which ought to have a
significant effect on the disc structure there.  There may be a
physical limit, beyond which the disc cannot respond, and accretion
will cease (see discussion of state 3, below).  This could possibly
lead to a time-dependent process \citep[e.g.,][]{spruittaam93},
perhaps analogous to the simulations of (e.g.)  \citet[][though their
stellar field does not connect outside $\rco$]{goodson3ea99}, and in
which there may still be a time-averaged net torque of zero.

          \subsubsection{State 3: $\rt > \rco$} \label{sub_state3}


Since no accretion onto the star occurs, state 3 may characterize the
non-accreting, weak-line T Tauri phase of pre-main sequence stellar
evolution.  Also, since the disc does not extend inside $\rco$, the
star feels no spin-up torques, only spin-down torques, and so it
cannot exist in spin equilibrium \citep{sunyaevshakura77b}.  A
possible magnetic field configuration of this state is illustrated in
the bottom panel of Figure \ref{fig_states}.

It is not yet clear under which conditions a system will be in this
state, though the relative values of differential torques (or
stresses) in the disc are certainly important.  Some authors
\citep[e.g.,][]{wang95, clarkeea95} have speculated that state 3
occurs when $|d \tmag / d \tacc|$ becomes greater than one
(\citealp{cameroncampbell93} suggested a value of 2) anywhere outside
$\rco$.  In this work, we have assumed that the disc will be
structured such that equation \ref{eqn_dtint} is satisfied \citep[see
\S \ref{sub_torques} and][]{rappaport3ea04}.  However, this assumption
must eventually break down for large enough $f$, large enough $\mu$,
or small enough $\mdotacc$.

The general question of what conditions govern a system in state 3, to
our knowledge, remains an unanswered astrophysical problem.  It is not
clear what will determine the location of $\rt$ in state 3 (since
neither of equations \ref{eqn_rt} and \ref{eqn_rt2} is then valid).
In addition to magnetic torques, outflows and/or radiation from the
star may be important \citep{johnstone95}.  Understanding this state
is probably relevant to understanding the transition from classical to
weak-line T Tauri phases, and it may even have further implications
for gas giant planet formation/migration \citep{lin3ea96,
trilling3ea02} and for the ultimate dissipation of the gas disc.

     \subsection{Conclusions} \label{sub_conclusions}

We have considered that the opening of magnetic field lines expected
from differential rotation in the star-disc interaction results in a
largely open field topology.  This significantly alters the torque
that a star receives from its accretion disc, compared to previous
models that assume a closed field.  Our main conclusions from this
work are the following:
\begin{enumerate}

\item{This more open field topology resuts in a weaker spin-down
torque felt by the star from the disc (\S \ref{sec_open}).  The
strongest possible torque occurs for intermediate magnetic coupling to
the disc.  Stronger coupling, as expected near the inner edge of the
disc, results in a spin-down torque that is more than an order of
magnitude below the torque found for the closed field assumption.}

\item{In the disc-locked, spin equilibrium state, this results in a
stellar spin rate that is much faster than predicted by previous
models (\S \ref{sub_spineq}).}

\item{We have identified and discussed three possible magnetic field
configurations in magnetic star-disc systems (\S \ref{sub_3states}).
The three configurations could represent, for example, an evolutionary
sequence for a system with a gradually decreasing mass accretion rate
(or, e.g., a gradually increasing stellar spin rate).  Our conclusions
about each state, in the context of T Tauri stars, are the following:
\begin{enumerate}

\item{Assuming strong magnetic coupling to the disc, slowly rotating
CTTS's should be in state 1 if $\mdotacc \ga 5 \times 10^{-9} M_\odot$
yr$^{-1}$ (eq.\ \ref{eqn_mtrans} for $B_* = 200$ G).  Since typical
accretion rates are higher than this, state 1 may represent a common
configuration in these systems.  In this state, the star feels no
spin-down torques from the disc.  However, if spin-down torques from
(e.g.) a stellar wind are significant, stars may be in spin
equilibrium in state 1, though they should not then be considered
`disc locked.'}

\item{State 2 is the typical configuration assumed in star-disc
interaction models.  For strong magnetic coupling in the disc, or if
stellar or disc winds are important, we find that a system can only be
in state 2 under special circumstances.  In particular, the accretion
rate must be lower than for state 1.}

\item{State 3 likely represents the non-accreting, weak-line T Tauri
phase.  Given that the accretion disc can restructure itself in
response to (e.g.) external magnetic torques, it it not yet clear when
a system will transition into this state.  This important evolutionary
phase requires more theoretical study.}

\end{enumerate}}

\item{These considerations, and additional issues from the literature,
suggest that slowly rotating CTTS's probably cannot be explained by a
disc-locking scenario (\S \ref{sub_problems}).}

\item{If slowly rotating CTTS's are in spin equilibrium, then another
spin down torque must be active in the system.  We suggest that this
might arise from magnetized stellar winds.}

\end{enumerate}

\section*{acknowledgements}

This work has been significanly influenced by open communication lines
with Dmitri Uzdensky, Keivan Stassun, and Arieh K\"onigl and
discussion with Cathie Clarke and Bob Mathieu, and we are grateful for
their contributions.  This research was supported by the National
Science and Engineering Research Council (NSERC) of Canada, McMaster
University, and the Canadian Institute for Theoretical Astrophysics
through a CITA National Fellowship.




\label{lastpage}
\end{document}